\documentclass[english]{article}
\usepackage[T1]{fontenc}
\usepackage[latin9]{inputenc}
\usepackage{amsthm}
\usepackage{amsmath}
\usepackage{amssymb}
\usepackage{esint}
\usepackage{babel}

\makeatletter

\numberwithin{equation}{section}
\numberwithin{figure}{section}

\makeatother

\title{\Large Constraints and symmetry in mechanics of affine motion}

\author{J. J. S\l awianowski, B. Go\l ubowska, and V. Kovalchuk\\
\\
Institute of Fundamental Technological Research,\\
Polish Academy of Sciences,\\
$5^{\rm B}$, Pawi\'{n}skiego str., 02-106 Warsaw, Poland\\
\\
{\it e-mails: jslawian@ippt.gov.pl,} \\
{\it bgolub@ippt.gov.pl, vkoval@ippt.gov.pl}}

\begin{document}

\maketitle

\begin{abstract}
The aim of this paper is to perform a deeper geometric analysis of
problems appearing in dynamics of affinely rigid bodies. First of
all we present a geometric interpretation of the polar and two-polar
decomposition of affine motion. Later on some additional constraints
imposed on the affine motion are reviewed, both holonomic and non-holonomic.
In particular, we concentrate on certain natural non-holonomic models
of the rotation-less motion. We discuss both the usual d'Alembert
model and the vakonomic dynamics. The resulting equations are quite
different. It is not yet clear which model is practically better.
In any case they both are different from the holonomic constraints
defining the rotation-less motion as a time-dependent family of symmetric
matrices of placements. The latter model seems to be non-geometric
and non-physical. Nevertheless, there are certain relationships between
our non-holonomic models and the polar decomposition.

\end{abstract}

\section{Affine constraints, geometry of the polar and two-polar decompositions}

Let us begin with a short review of our earlier results concerning
the mechanics of affinely-rigid body \cite{JJS_74,JJS_75_1,JJS_75_2}.
To be honest, some of them are also partially contained in Eringen's
theory of micromorphic media, i.e., continua of infinitesimal affine
bodies \cite{Erin_68}. Later on, we developed the theory in various
aspects \cite{Gol_01,Gol_02,Gol_03,Kov_10,Mart_JJS_10,Rozko_05,Rozko_10,JJS_82_1,JJS_82_2,JJS_87,JJS_88,JJS_04,JJS_05,JJS-Kov_03,JJS-Kov_04,JJSall_04,JJSall_05,JJSall_10,JJSall_11,JJSall_12,JJS-AS_93,AT-JJS_90}
and some of our results were confirmed and developed by many people
\cite{Reilly_96,Reilly_98,Papa_01,Rubin_85,Rubin_86,Sol-Papa_99,Sol-Papa_00,Sousa_94}.
Let us also mention the papers like \cite{Burov_96,Capriz_89,Cheva_04,Coh-Mun_89,Rob-Wulf-Lamb_02,Wulf-Rob_02}.
Nevertheless, in spite of numerous applications the topic does not
belong to commonly known standards, and because of this a brief repetition
seems to be necessary.

Let us consider a system of material points moving in $n$-dimensional
physical space $M$; we assume $M$ to be an affine space with the
linear space of translations $V$, endowed also with the symmetric
and positively-definite metric tensor $g\in V^{*}\otimes V^{*}$.
If necessary, the translation vector from $x\in M$ to $y\in M$ will
be denoted by $\overrightarrow{xy}$. The material space, i.e., the
set of material points will be also an affine space $N$ of the same
dimension $n$, with the linear space of translations $U$. The material
metric tensor will be denoted by $\eta\in U^{*}\otimes U^{*}$, and
translations vectors by $\overrightarrow{ab}$ for $a,b\in U$. As
usual, we say that a mapping $\phi:N\rightarrow M$ is affine if it
preserves all affine relationships, i.e., there exists a linear mapping
$L\left[\phi\right]:U\rightarrow V$, denoted also as $D\phi\in L\left(U,V\right)$
such that 
\begin{equation}
\overrightarrow{\phi\left(a\right)\phi\left(b\right)}=
L\left[\phi\right]\overrightarrow{ab}\label{eq:1}
\end{equation}
for any pair of material points, $a,b\in N$. If $y^{i}$, $a^{K}$
are affine coordinates respectively in $M$ and $N$, this means obviously
that $\phi$ is analytically given by first-order polynomials:
\begin{equation}
y^{i}=x^{i}+\varphi^{i}\!_{K}a^{K}.\label{eq:2}
\end{equation}

Obviously, this definition is valid for any, not necessarily equal
dimensions of $N$, $M$. The set of all affine mappings of $N$ onto
$M$ will be denoted by $Aff\left(N,M\right)$, and the set of all
one-to-one affine mappings of $N$ onto $M$ is denoted by $AffI\left(N,M\right)$
(affine isomorphisms). Obviously, $AffI\left(N,M\right)$ is non-empty
only if $\dim N=\dim M$, and for any $\phi\in AffI\left(N,M\right)$, $\varphi=L\left[\phi\right]\in LI\left(U,V\right)$, i.e., it is
a linear isomorphism of $U$ onto $V$. The groups of affine and linear
isomorphisms of $M$ and $V$ will be denoted by $GAff\left(M\right)$, $GL\left(V\right)$. They are open subsets of $Aff\left(M\right)$,
$L\left(V\right)$, i.e., of the sets of all affine and linear mappings
of $M$ and $V$ into themselves. 

Every choice of affine coordinates $a^{K}$, $y^{i}$ in $N$, $M$
pre-assumes two things: a choice the origins $\mathfrak{O}\in N$,
$\mathfrak{o}\in M$ of coordinates in $N$, $M$ and a choice of
bases $\left(\ldots,E_{A},\ldots\right)$, $\left(\ldots,e_{i},\ldots\right)$
in $U$, $V$, or equivalently, a choice of dual bases $\left(\ldots,E^{A},\ldots\right)$,
$\left(\ldots,e^{i},\ldots\right)$ in $U^{*}$, $V^{*}$. Then we
have 
\begin{equation}
a^{K}(P)=\left\langle E^{K},\overrightarrow{\mathfrak{O}P}\right\rangle \;,\quad y^{i}(p)=\left\langle e^{i},\overrightarrow{\mathfrak{o}p}\right\rangle \label{eq:3}
\end{equation}
for any points $P\in N$, $p\in M$. When the constant co-moving mass
distribution in $N$ is fixed and described by positive measure $\mu$
on $N$, then it is natural to choose $\mathfrak{O}\in N$ as the
centre of mass, 
\begin{equation}
\int\overrightarrow{\mathfrak{O}P}\,d\mu=0.\label{eq:4}
\end{equation}

The point $\mathfrak{O}$ is uniquely defined when $m=\mu(N)$ is
finite, what is physically always assumed. With such a choice of $\mathfrak{O}$,
the quantities $x^{i}$ in (\ref{eq:2}) are the current coordinates
of the centre of mass in $M$, $\mathfrak{o}_{\phi}=\phi(\mathfrak{O})$.
Let us stress that for any, not necessarily affine, configuration
$\mathfrak{o}_{\phi}$ is defined by the condition 
\begin{equation}
\int\overrightarrow{\mathfrak{o}_{\phi}p}\,d\mu_{\phi}(p)=0,\label{eq:5}
\end{equation}
where $\mu_{\phi}$ denotes the $\phi$-transport of the measure $\mu$
from $N$ to $M$. The equality $\mathfrak{o}_{\phi}=\phi\left(\mathfrak{O}\right)$
holds only for affine configurations. 

When the choice of $\mathfrak{O}$ is fixed as above, then the configuration
space of affinely-rigid body, i.e., the manifold of affine isomorphisms
of $N$ onto $M$, $AffI\left(N,M\right)$ becomes canonically identified
with the Cartesian product $M\times LI\left(U,V\right)$: 
\begin{equation}
\phi\equiv\left(\phi\left(\mathfrak{O}\right),L\left[\phi\right]\right)=
\left(\dots,x^{i},\dots;\dots,\varphi^{i}\!_{K},\dots\right).\label{eq:6}
\end{equation}
This is the splitting of degrees of freedom into translational and
internal ones.

All those concepts are purely affine and the metric tensors $g$,
$\eta$ occur only on the dynamical level. Let us mention, there are
also purely affine, metric-free dynamical models \cite{JJSall_04,JJSall_05}, but it is quite a different story.

The affine groups $GAff\left(M\right)$, $GAff\left(N\right)$ act
on the configuration space of affine body $AffI\left(N,M\right)$
through the left and right superpositions. Namely, any $\left(\mathcal{A},\mathcal{B}\right)\in GAff\left(M\right)\times GAff\left(N\right)$
transforms the configuration $\phi\in AffI\left(N,M\right)$ as follows:
\begin{equation}
\phi\rightarrow\mathcal{A}\circ\phi\circ\mathcal{B}.\label{eq:7}
\end{equation}
Obviously, the action of $GAff\left(M\right)$ does commute with that
of $GAff\left(N\right)$. They are respectively the spatial and material
transformation groups. 

Describing affine configurations as in (\ref{eq:6}) we can represent
the actions (\ref{eq:7}) of $\mathcal{A}\in GAff(M)$, $\mathcal{B}\in GAff(N)$
as follows: 
\begin{equation}
\mathcal{A}\in GAff(M):\quad\left(x,\varphi\right)\rightarrow\left(\mathcal{A}(x),
L\left[\mathcal{A}\right]\varphi\right),\label{eq:8}
\end{equation}
\begin{equation}
\mathcal{B}\in GAff(N):\quad\left(x,\varphi\right)\rightarrow\left(t\left[\varphi
\cdot\overrightarrow{\mathfrak{O}\mathcal{B}\left(\mathfrak{O}\right)}\right],
\varphi L\left[\mathcal{B}\right]\right),\label{eq:9}
\end{equation}
where for any $v\in V$, $u\in U$, the symbols $t[v]$, $t[u]$ denote
translation operations in $M$ and $N$, i.e., such affine transformations
of $M$ and $N$ that 
\begin{equation}
\overrightarrow{xt[v](x)}=v,\qquad\overrightarrow{at[u](a)}=u\label{eq:10}
\end{equation}
for any $x\in M$, $a\in N$. If, after the material origin $\mathfrak{O}\in N$
is fixed, $\mathcal{B}$ is identified with $\left(B,b\right)\in GL\left(U\right)\underset{\sim}{\times}U$
(semi-direct product), then (\ref{eq:9}) becomes 
\begin{equation}
\left(B,b\right):\quad\left(x,\varphi\right)\rightarrow\left(t\left[\varphi b\right]\left(x\right),\varphi B\right).\label{eq:11}
\end{equation}
Analytically, (\ref{eq:8}) and (\ref{eq:9})/(\ref{eq:11}) are respectively
given by
\begin{equation}
\left(\ldots,x^{i}.\ldots;\ldots\varphi^{j}\!_{K},\ldots\right)
\rightarrow\left(\ldots,A^{i}\!_{m}x^{m}+a^{i},\ldots;\ldots A^{j}\!_{m}\varphi^{m}\!_{K},\ldots\right),\label{eq:12}
\end{equation}
\begin{equation}
\left(\ldots,x^{i}.\ldots;\ldots\varphi^{j}\!_{K},\ldots\right)
\rightarrow\left(\ldots,x^{i}+\varphi^{i}\!_{M}x^{M},\ldots;\ldots
\varphi^{j}\!_{L}B^{L}\!_{K},\ldots\right).\label{eq:13}
\end{equation}
The structural difference between spatial (Eulerian) and material
(Lagrangian) transformations is easily seen here. Let us observe that
$GL(V)$, $GL(U)$ act also on the manifold $Q_{int}=LI(U,V)$ of internal/relative degrees of freedom through the obvious formulas:
\begin{equation}
\varphi\rightarrow A\varphi B\;,\qquad\left(A,B\right)\in GL(V)\times GL(U).\label{eq:14}
\end{equation}
Obviously, this action is non-effective and corresponding kernel is
given by the subgroup:
\begin{equation}
\left\{ \left(\lambda Id_{V},\lambda^{-1}Id_{U}\right):\lambda\in\mathbb{R}\setminus\left\{ 0\right\} \right\} \subset GL(V)\times GL(U).\label{eq:15}
\end{equation}
The subgroups of $GL(V)$, $GL(U)$ and those of $GAff(M)$, $GAff(N)$
act in a natural way on the internal and total configuration spaces
$Q_{int}=LI(U,V)$, $Q=Aff(N,M)\simeq M\times Q_{int}$. Let us mention
a few most important of them: orthogonal groups $O(V,g)$, $O(U,\eta)$
their rotation subgroups $SO(V,g)$, $SO(U,\eta)$, special linear
groups $SL(V)$, $SL(U)$ or one-dimensional dilatation subgroups
$Dil(V)=\left\{ \lambda Id_{V}:\lambda\in\mathbb{R}\setminus\left\{ 0\right\} \right\} $,
$Dil(U)=\left\{ \lambda Id_{W}:\lambda\in\mathbb{R}\setminus\left\{ 0\right\} \right\} $.
In the total configuration space $Q$, when translational degrees
of freedom are taken into account, those groups are semi-directly
extended by translations $T(M)\simeq V$, $T(N)\simeq U$ to the corresponding
affine subgroups: Euclidean $E\left(M,g\right)$, $E\left(N,\eta\right)$,
isochoric $SAff(M)$, $SAff(N)$ and dilatations/translations $Dil(M)$,
$Dil(N)$. The meaning of symbols is obvious. Let us only remind a
few definitions. $A\in O(V,g)$, $B\in O(U,\eta)$, $\varphi\in O(U,\eta;V,g)$
when they preserve the metric tensors, thus, 
\begin{equation}
g_{ij}=g_{kl}A^{k}\!_{i}A^{l}\!_{j},\quad\eta_{AB}=
\eta_{CD}B^{C}\!_{A}B^{D}\!_{B},\quad\eta_{AB}=
g_{ij}\varphi^{i}\!_{A}\varphi^{j}\!_{B}.\label{eq:16}
\end{equation}
$A\in SO(V,g)$, $B\in SO(U,\eta)$, when not only $\left|\det A\right|=\left|\det B\right|=1$, but just $\det A=\det B=1$. When orientations $\rho$, $\omega$ in $U$, $V$ are fixed and $\det\varphi=1$ in some orthonormal positively oriented
bases in $U$, $V$, then we say that $\varphi\in SO\left(U,\eta,\rho;V,g,\omega\right)$ when $\det\varphi=1$ and that $\varphi\in O\left(U,\eta;V,g\right)$ when $\left|\det\varphi\right|=1$ Similarly, we say that $A\in SL(V)$, $B\in SL(U)$ when $\det A=\det B=1$ but without orthogonality condition (\ref{eq:16}). And similarly $\varphi\in SL\left(U,\rho;V,\omega\right)$ when $\det\varphi=1$ in some positively oriented bases and that $\varphi\in UL(U,V)$ when $\left|\det\varphi\right|=1$. If $\left|\det A\right|=\left|\det B\right|=1$ we say that, $A$, $B$ are unimodular and write that $A\in UL(V)$, $B\in UL(U)$. If $\det A=\det B=1$ we say that they are special linear. 

Lie algebras of $GL^{+}(V)$, $GL^{+}(U)$, the proper (positive-determinants)
subgroups of $GL(V)$, $GL(U)$ are isomorphic with the commutator Lie
algebras of all linear mappings $GL(V)^{\prime}\simeq L(V)$, $GL(U)^{\prime}\simeq L(U)$. And Lie algebras $SO(V,g)^{\prime}$, $SO(U\eta)^{\prime}$ consist respectively of $g$- and $\eta$-skew-symmetric elements of $L(V)$, $L(U)$: 
\begin{equation}
a^{i}\!_{j}=-a_{j}\!^{i}=-g_{jk}g^{il}a^{k}\!_{l},\quad b^{A}\!_{B}=-b_{B}\!^{A}=-\eta_{BC}\eta^{AD}b^{C}\!_{D}.\label{eq:17}
\end{equation}
Lie algebras $SL(V)^{\prime},SL(U)^{\prime}$ consist of trace-less
linear mappings:
\begin{equation}
Tr\; a=a^{i}\!_{i}=0,\quad Tr\; b=b^{K}\!_{K}=0.\label{eq:18}
\end{equation}
As usual, when dealing with group-theoretic degrees of freedom, it
is convenient to use non-holonomic Lie-algebraic velocities.
In the case of affine systems we shall use the term ``affine velocity'';
Eringen referred to them as ``gyrations''. The spatial and material
affine velocities of internal motion are given by 
\begin{equation}
\Omega=\frac{d\varphi}{dt}\varphi^{-1}\in L(V),\quad\widehat{\Omega}=\varphi^{-1}\frac{d\varphi}{dt}=
\varphi^{-1}\Omega\varphi\in L(U).\label{eq:19}
\end{equation}
Besides the usual velocity of translational motion, $v^{i}=dx^{i}/dt$,
one uses also its co-moving representation:
\begin{equation}
\widehat{v}=\varphi^{-1}v,\quad\widehat{v}^{A}=
\left(\varphi^{-1}\right)^{A}\!_{i}v^{i}.\label{eq:20}
\end{equation}
When gyroscopic constraints of metrically-rigid motion are imposed,
\begin{equation}
\varphi\in SO(U,\eta,\rho;V,g,\omega), 
\end{equation}
then $\Omega$, $\widehat{\Omega}$ are respectively $g$- and $\eta$-skew-symmetric, i.e., they satisfy (\ref{eq:17}) when substituted instead $a$, $b$. This is the alternative, ``anholonomic'' representation of those holonomic constraints.

When the body is incompressible, $\det\varphi=1$, then $\Omega$, $\widehat{\Omega}$ are trace-less, i.e., they satisfy (\ref{eq:18}).

Gyroscopic and isochoric constraints are, obviously, holonomic. Geometrically
this has to do with the fact that their affine velocities are elements
of the commutator Lie subalgebras of $L(V)$, $L(U)$. There are, however,
another interesting cases of non-holonomic constraints of spatially
and materially rotation-less motion. In the first case $\Omega$ is
$g$-symmetric, in the second one $\widehat{\Omega}$ is $\eta$-symmetric, i.e.,
respectively, 
\begin{equation}
\Omega^{i}\!_{j}=\Omega_{j}\!^{i}=g_{jk}g^{il}\Omega^{k}\!_{l},
\quad\widehat{\Omega}^{A}\!_{B}=\widehat{\Omega}_{B}\!^{A}=
\eta_{BC}\eta^{AD}\widehat{\Omega}^{C}\!_{D}.\label{eq:21}
\end{equation}
Let us observe that unlike in the holonomic gyroscopic constraints,
the two conditions (\ref{eq:21}) are non-equivalent and describe
different non-holonomic constraints. Namely, the first, i.e., spatial,
condition in (\ref{eq:21}) is materially represented by
\begin{equation}
\widehat{\Omega}^{A}\!_{B}=G_{BC}G^{AD}\widehat{\Omega}^{C}\!_{D},\label{eq:22}
\end{equation}
where $G_{KL}$ are components the Green deformation tensor, 
\begin{equation}
G_{KL}=g_{ij}\varphi^{i}\!_{K}\varphi^{j}\!_{L},\label{eq:23}
\end{equation}
and $G^{KL}$ represent its contravariant inverse, 
\begin{equation}
G^{KM}G_{ML}=\delta^{K}\!_{L},\quad G^{KL}\neq\eta^{KM}\eta^{LN}G_{NM}.\label{eq:24}
\end{equation}
Similarly, the second, i.e., material, equation in (\ref{eq:21}) is
in the spatial language equivalent to 
\begin{equation}
\Omega^{i}\!_{j}=C^{^{ik}}C_{jm}\Omega^{m}\!_{k},\label{eq:25}
\end{equation}
where $C_{ij}$ are components of the Cauchy deformation tensor, 
\begin{equation}
C_{ij}=\eta_{AB}\left(\varphi^{-1}\right)^{A}\!_{i}\left(
\varphi^{-1}\right)^{B}\!_{j},\label{eq:26}
\end{equation}
and $C^{ij}$ are coordinates of its contravariant inverse: 
\begin{equation}
C^{ik}C_{kj}=\delta^{i}\!_{j},\quad C^{ij}\neq g^{ik}g^{jl}C_{kl}.
\label{eq:27}
\end{equation}
The inequalities in (\ref{eq:24}), (\ref{eq:27}) show that the thoughtless
use of the kernel-index convention may be misleading. Having in $U$, $V$
two metric-like tensors $G$, $\eta$ and $C$, $g$ one can construct
two mixed tensors: 
\begin{equation}
\widehat{G}^{A}\!_{B}:=\eta^{AC}G_{CB},\quad\widehat{C}^{i}\!_{j}:=
g^{ik}C_{kj}\label{eq:28}
\end{equation}
and the family of scalars, e.g., 
\begin{equation}
I_{k}=Tr\left(\widehat{G}^{k}\right)=Tr\left(\widehat{C}^{-k}\right).
\label{eq:29}
\end{equation}
Those scalars are invariant under the action (\ref{eq:14}) of the
subgroup $O(V,g)\times O(U,\eta)\subset GL(V)\times GL(U)$. They
are basic orthogonal deformation invariants of $\varphi$. According
to the Cayley-Hamilton theorem, there are only $n$ independent invariants,
e.g., those corresponding to $k=1,\dots,n$; any other invariant is
their function.

Deformation invariants tell us how strongly the body is stretched/contracted, but they do not contain any information about the spatial
or material orientation of the stretching. This information is encoded
in directions of the main axes of deformation tensors $G$, $C$. More
precisely, let $L_{a}[\varphi]$, $R_{a}[\varphi]$, $a=1,\dots,n$,
be orthonormal basic eigenvectors of $\widehat{C}[\varphi]$, $\widehat{G}[\varphi]$:
\begin{eqnarray}
&&\widehat{C}L_{a}=\lambda_{a}^{-1}L_{a},\quad\widehat{G}R_{a}=
\lambda_{a}R_{a},\label{eq:30a} \\
&&g\left(L_{a},L_{b}\right)=g_{ij}L^{i}\!_{a}L^{j}\!_{b}=\delta_{ab}=
\eta_{CD}R^{C}\!_{a}R^{D}\!_{b}=\eta\left(R_{a},R_{b}\right).\label{eq:30b}
\end{eqnarray}
Their dual covectors $L^{a}[\varphi]\in V^{*}$, $R^{a}[\varphi]\in U^{*}$
satisfy
\begin{eqnarray}
C[\varphi]&= & \underset{a}{\sum}\lambda_{a}^{-1}[\varphi]L^{a}[\varphi]\otimes L^{a}[\varphi],\label{eq:31a} \\
G[\varphi]&= & \underset{a}{\sum}\lambda_{a}[\varphi]R^{a}[\varphi]\otimes R^{a}[\varphi].\label{eq:31b}
\end{eqnarray}
It is convenient to introduce the symbols $Q^{a}$, $q^{a}$,
\begin{equation}
Q^{a}=\exp\left(q^{a}\right)=\sqrt{\lambda_{a}}.\label{eq:32}
\end{equation}

The ordered bases $L=\left(\dots,L_{a},\dots\right)$, $R=\left(\dots,R_{a},\dots\right)$
are identified with isomorphisms $L:\mathbb{R}^{n}\rightarrow V$,
$R:\mathbb{R}^{n}\rightarrow U$ and their dual co-bases $L^{-1}=\left(\dots,L^{a},\dots\right)$,
$R^{-1}=\left(\dots,R^{a},\dots\right)$ may be interpreted as the
inverse isomorphisms $L^{-1}:V\rightarrow\mathbb{R}^{n}$, $R^{-1}:U\rightarrow\mathbb{R}^{n}$.
The diagonal matrix with diagonal entries $Q^{a}$, $Diag\left(\dots,Q^{a},\dots\right)$,
is identified with an isomorphism $D:\mathbb{R}^{n}\rightarrow\mathbb{R}^{n}$.
And finally, the isomorphism $\varphi:U\rightarrow V$ may be represented
as: 
\begin{equation}
\varphi=LDR^{-1}.\label{eq:33}
\end{equation}

In this way affine configurations $\varphi$ are identified with the
triplets consisting of two gyroscopic configurations of metrically-rigid bodies $L$, $R$, and of the system of $n$ material points on
$\mathbb{R}$ (deformation invariants).

For any pair of linear bases, e.g., $\left(\dots,R_{a}[\varphi],\dots\right)$
in $U$ and $\left(\dots,L_{a}[\varphi],\dots\right)$ in $V$, there
exists exactly one linear mappings $U[\varphi]$ of $U$ onto $V$
such that 
\begin{equation}
L_{a}[\varphi]=U[\varphi]R_{a}[\varphi],\quad a=1,\dots,n.\label{eq:34}
\end{equation}
When the bases are respectively $\eta$- and $g$-orthonormal, then
$U[\varphi]\in O(U,\eta;V,g)$, and $\varphi$ may be expressed as
\begin{equation}
\varphi=U[\varphi]A[\varphi]=B[\varphi]U[\varphi],\label{eq:35}
\end{equation}
where the linear mappings $A[\varphi]\in GL(U)$, $B[\varphi]\in GL(V)$
are symmetric in the $\eta$- and $g$-sense and positively-definite.
Obviously, by the positive definiteness we mean 
\begin{equation}
\eta\left(A\left[\varphi\right]z,z\right)>0,\quad g\left(B\left[\varphi\right]\omega,\omega\right)>0\label{eq:36}
\end{equation}
and by the symmetry 
\begin{equation}
\eta\left(A\left[\varphi\right]u,v\right)=\eta\left(u,A\left[\varphi\right]
v\right),\quad g\left(B\left[\varphi\right]x,y\right)=g\left(x,B\left[\varphi\right]y\right)
\label{eq:37}
\end{equation}
for non-vanishing $z\in U$, $\omega\in V$ and for any $u,v\in U$, $x,y\in V$.

Unlike the two-polar splitting (\ref{eq:33}) the both versions of
the polar splitting (\ref{eq:35}) are unique. And, obviously, $A[\varphi]$, $B[\varphi]$ are related to each other by the $U[\varphi]$-similarity :
\begin{equation}
A[\varphi]=U[\varphi]^{-1}B[\varphi]U[\varphi].\label{eq:38}
\end{equation}
Analytically, in the matrix language $L$, $R$, $U$ are orthogonal,
$D$ is diagonal positive, and $A$, $B$ are symmetric positively definite
matrices.

Transformations (\ref{eq:7}), (\ref{eq:14}) act on affine velocities
according to the obvious rules:
\begin{equation}
\Omega\rightarrow A\Omega A^{-1},\quad\widehat{\Omega}\rightarrow B^{-1}\widehat{\Omega}B.\label{eq:39}
\end{equation}
According to the same rules orthogonal transformations act on the $g$- and
$\eta$-skew-symmetric angular velocities.

In analogy to non-holonomic affine velocities, one introduces their
dual affine spin quantities,
\begin{equation}
\Sigma=\varphi P,\quad\widehat{\Sigma}=P\varphi=\varphi^{-1}\Sigma\varphi,\label{eq:40}
\end{equation}
where $P$ denotes the system of canonical momenta $P^{A}\!_{i}$
conjugate to $\varphi^{i}\!_{A}$, $\Sigma^{i}\!_{j}$, $\widehat{\Sigma}^{A}\!_{B}$
are momentum mappings of the transformation group (\ref{eq:14}).
In other words, they are Hamiltonian generators of this group. Their
Poisson brackets correspond to the structure constants of $GL(V)$, $GL(U)$:
\begin{eqnarray}
\left\{ \Sigma^{i}\!_{j},\Sigma^{k}\!_{l}\right\}  & = & \delta^{i}\!_{l}\Sigma^{k}\!_{j}-\delta^{k}\!_{j}\Sigma^{i}\!_{l},
\label{eq:41a}\\
\left\{ \widehat{\Sigma}^{A}\!_{B},\widehat{\Sigma}^{C}\!_{D}\right\}  & = & \delta^{C}\!_{B}\widehat{\Sigma}^{A}\!_{D}-\delta^{A}\!_{D}
\widehat{\Sigma}^{C}\!_{B},\label{eq:41b}\\
\left\{ \Sigma^{i}\!_{j},\widehat{\Sigma}^{A}\!_{B}\right\}  & = & 0.
\label{eq:41c}
\end{eqnarray}
And, obviously, the following holds: 
\begin{equation}
\left\{ \Sigma^{i}\!_{j},\varphi^{k}\!_{A}\right\} =\delta^{k}\!_{j}\varphi^{i}\!_{A},\quad\left\{ \widehat{\Sigma}^{A}\!_{B},\varphi^{k}\!_{C}\right\} =\delta^{A}\!_{C}\varphi^{k}\!_{B}.\label{eq:42}
\end{equation}
Transformations (\ref{eq:7}), (\ref{eq:14}) act on $\Sigma^{i}\!_{j}$, $\widehat{\Sigma}^{A}\!_{B}$
just like on $\Omega$, $\widehat{\Omega}$ (\ref{eq:39}): 
\begin{equation}
\Sigma\rightarrow A\Sigma A^{-1},\quad\widehat{\Sigma}\rightarrow B^{-1}\widehat{\Sigma}B.\label{eq:43}
\end{equation}
The same may be done with the gyroscopic degrees of freedom of the
two-polar and polar decompositions. The $\mathbb{R}^{n}$-comoving
angular velocity $\widehat{\chi}^{a}\!_{b}$ of the $L$-top and the
$V$-spatial representation $\chi^{i}\!_j$ are given by
\begin{eqnarray}
\widehat{\chi}^{a}\!_{b}=\left\langle L^{a},\frac{dL_{b}}{dt}\right\rangle =L^{a}\!_{i}\frac{dL^{i}\!_{b}}{dt},\label{eq:44a} \\
\chi^{i}\!_{j}=\frac{dL^{i}\!_{a}}{dt}L^{a}\!_{j},\quad\chi=\widehat{\chi}^{a}\!_{b}L_{a}\otimes L^{b}.\label{eq:44b}
\end{eqnarray}
And similarly, the $\mathbb{R}^{n}$-comoving and $U$-spatial components
of the angular velocity of the $R$-top, $\widehat{\vartheta}^{a}\!_{b}$, $\vartheta^{K}\!_{L}$
are given by
\begin{eqnarray}
\widehat{\vartheta}^{a}\!_{b}=\left\langle R^{a},\frac{dR_{b}}{dt}\right\rangle =R^{a}\!_{K}\frac{dR^{K}\!_{b}}{dt},\label{eq:45a} \\
\vartheta^{K}\!_{L}=\frac{dR^{K}\!_{a}}{dt}R^{a}\!_{L},\quad\vartheta=
\widehat{\vartheta}^{a}\!_{b}R_{a}\otimes R^{b}.\label{eq:45b}
\end{eqnarray}
Obviously, the ``comoving'' angular velocities $\widehat{\chi}^{a}\!_{b}$, $\widehat{\vartheta}^{a}\!_{b}$
are $\delta$-antisymmetric, while the ``spatial'' ones, $\chi^{i}\!_{j}$, $\vartheta^{K}\!_{L}$,
are respectively $g$- and $\eta$-antisymmetric. The
dual spins conjugate to $\widehat{\chi}^{a}\!_{b}$, $\widehat{\vartheta}^{a}\!_{b}$,
$\chi^{i}\!_{j}$, $\vartheta^{K}\!_{L}$ are also skew-symmetric matrices,
denoted respectively by $\widehat{\rho}^{a}\!_{b}$, $\widehat{\tau}^{a}\!_{b}$,
$\rho^{i}\!_{j}$, $\tau^{K}\!_{L}$. When canonical momenta conjugate
to $q^{a}$ are denoted by $p_{a}$, then the duality relations have
the form:
\begin{eqnarray}
\left\langle \left(\widehat{\rho},\widehat{\tau},p\right),\left(\widehat{\chi},
\widehat{\vartheta},\dot{q}\right)\right\rangle  & = & \left\langle \left(\rho,\tau,p\right),\left(\chi,\vartheta,\dot{q}\right)\right\rangle \nonumber \\
& = & p_{a}\dot{q}^{a}+\frac{1}{2}Tr\left(\widehat{\rho}\widehat{\chi}\right)+
\frac{1}{2}Tr\left(\widehat{\tau}\widehat{\vartheta}\right)\nonumber\\
& = & p_{a}\dot{q}^{a}+\frac{1}{2}Tr\left(\rho\chi\right)+\frac{1}{2}Tr\left(
\tau\vartheta\right).\label{eq:46}
\end{eqnarray}
It is clear that $\rho$, $\tau$ are Hamiltonian generators of the
transformation groups, 
\begin{equation}
\varphi\rightarrow A\varphi B^{-1},\qquad A\in SO(V,g),\quad B\in SO(U,\eta),\label{eq:47}
\end{equation}
so, they are equal respectively to the spin and minus vorticity. Let
us remind that the spin and vorticity are doubled $g$- and 
$\eta$-skew-symmetric parts of $\Sigma^{i}\!_{j}$, 
$\widehat{\Sigma}^{A}\!_{B}$. When we use the polar splitting (\ref{eq:35}), then the gyroscopic $U$-motion is characterized by angular velocity in the co-moving and spatial representations, respectively: 
\begin{equation}
\widehat{\omega}=U^{-1}\frac{dU}{dt},\quad\omega=\frac{dU}{dt}U^{-1}=
U\widehat{\omega}U^{-1}.\label{eq:48}
\end{equation}

\section{Kinetic energy, equations of motion, additional constraints}

It may be easily shown that the kinetic energy of the classical affine
body is given by 
\begin{equation}
T=T_{tr}+T_{int}=\frac{m}{2}g_{ij}\frac{dx^{i}}{dt}\frac{dx^{j}}{dt}+
\frac{1}{2}g_{ij}\frac{d\varphi^{i}\!_{A}}{dt}\frac{d\varphi^{j}\!_{B}}
{dt}J^{AB},\label{eq:49}
\end{equation}
where $m\in\mathbb{R}$ and $J\in U\otimes U$ are constant inertial
parameters of affine degrees of freedom, 
\begin{equation}
m=\int\limits_{N}d\mu(a),\qquad J^{AB}=\int\limits_{N}a^{A}a^{B}d\mu(a).\label{eq:50}
\end{equation}
Therefore, $m$ is the total mass of the body and $J^{AB}$ is the
quadrupole momentum of the mass distribution, algebraically equivalent
to the co-moving inertial tensor. Let us repeat that, following (\ref{eq:4}),
the dipole momentum vanishes: 
\begin{equation}
\int a^{A}d\mu(a)=0.\label{eq:51}
\end{equation}
And the higher multipoles, although non-vanishing, do not contribute
to the affine motion.

The kinetic energy (\ref{eq:49}) is invariant under the action (\ref{eq:14})
of $O(V,g)\times O(U,J^{-1})\subset GL(V)\times GL(U)$, and of course
under the action of translations. It is important that because of
the essential dependence of its metric tensor on $g$, $J$, it fails
to be invariant under the total action of affine groups $GAff(M)$, $GL(V)$, $GAff(N)$, $GL(U)$ .After substitution of the polar decomposition (\ref{eq:35})  to (\ref{eq:49}), it becomes 
\begin{eqnarray}
T_{int} & = & \frac{1}{2}\eta_{KL}\frac{dA^{K}\!_{A}}{dt}\frac{dA^{L}\!_{B}}{dt}J^{AB}
+\eta_{KL}\widehat{\omega}^{K}\!_{C}A^{C}\!_{A}\frac{dA^{L}\!_{B}}{dt}J^{AB}
\nonumber \\
& + & \frac{1}{2}\eta_{KL}\widehat{\omega}^{K}\!_{C}\widehat{\omega}^{L}\!_{D}A^{C}
\!_{A}A^{D}\!_{B}J^{AB}.\label{eq:52}
\end{eqnarray}
The first term represents the kinetic energy of deformative vibrations,
the second one is the Coriolis coupling between deformative and rotational
motion, and the third term describes the centrifugal coupling of rotations
and deformations. The lower-case indices of the third term are contracted
with the $A$-deformed inertial tensor, $A^{C}\!_{A}A^{D}\!_{B}J^{AB}$.
A similar formula holds for the second of the polar decompositions
(\ref{eq:35}). When we use the purely analytical language and orthonormal
coordinates, $\eta_{KL}=\delta_{KL}$, then the formula (\ref{eq:52})
may be written in the following brief matrix form: 
\begin{equation}
T_{int}=\frac{1}{2}Tr\left(J\left(\frac{dA}{dt}\right)^{2}\right)
+Tr\left(AJ\frac{dA}{dt}\widehat{\omega}\right)-\frac{1}{2}Tr\left(AJA
\widehat{\omega}^{2}\right).\label{eq:53}
\end{equation}
Substituting the two-polar decomposition (\ref{eq:33}) and its by-products (\ref{eq:44a}), (\ref{eq:44b}), (\ref{eq:45a}), (\ref{eq:45b}) to (\ref{eq:49}), we obtain a rather complicated formula:
\begin{eqnarray}
T_{int} & = & \frac{1}{2}Tr\left(\left(\frac{dD}{dt}\right)^{2}R^{-1}JR\right)\nonumber \\
& + & \frac{1}{2}Tr\left(\frac{dD}{dt}\widehat{\chi}DR^{-1}JR\right)-\frac{1}{2}
Tr\left(D\widehat{\chi}\frac{dD}{dt}R^{-1}JR\right)\nonumber \\
 & + & \frac{1}{2}Tr\left(\widehat{\vartheta}D\frac{dD}{dt}R^{-1}JR\right)
-\frac{1}{2}Tr\left(\frac{dD}{dt}D\widehat{\vartheta}R^{-1}JR\right)\nonumber\\
 & - & \frac{1}{2}Tr\left(D\widehat{\chi}^{2}DR^{-1}JR\right)-\frac{1}{2}
Tr\left(\widehat{\vartheta}D^{2}\widehat{\vartheta}R^{-1}JR\right)\nonumber \\
 & + & \frac{1}{2}Tr\left(\widehat{\vartheta}D\widehat{\chi}DR^{-1}JR\right)
+\frac{1}{2}Tr\left(D\widehat{\chi}D\widehat{\vartheta}R^{-1}JR\right).
\label{eq:54}
\end{eqnarray}
It is seen that the complication is due to the term $R^{-1}JR$. And indeed,
the to polar splitting is computationally optimal in the special case
of inertially isotropic body, when 
\begin{equation}
J^{AB}=I\eta^{AB}\label{eq:55}
\end{equation}
and the kinetic energy is invariant under the material orthogonal
group $O(U,\eta)$. Then 
\begin{equation}
\left(R^{-1}JR\right)^{ab}=I\delta^{ab}\label{eq:56}
\end{equation}
and the formula (\ref{eq:54}) simplifies to 
\begin{equation}
T_{int}=\frac{I}{2}Tr\left(\left(\frac{dD}{dt}\right)^{2}\right)
+ITr\left(D\widehat{\chi}D\widehat{\vartheta}\right)-\frac{I}{2}
Tr\left(D^{2}\widehat{\chi}^{2}\right)-\frac{I}{2}Tr\left(D^{2}
\widehat{\vartheta}^{2}\right).\label{eq:57}
\end{equation}
It is clear that combining appropriately $\widehat{\chi}$, $\widehat{\vartheta}$
we can avoid interference terms. This is explicitly seen when instead
of the usual kinetic formulas like (\ref{eq:49}), (\ref{eq:52}), (\ref{eq:53}),
(\ref{eq:57}) one uses their canonical forms based on the Legendre
transformation, like, e.g., 
\begin{equation}
p_{i}=mg_{ij}\frac{dx^{j}}{dt},\quad p^{A}\!_{i}=g_{ij}\frac{d\varphi^{j}\!_{B}}{dt}J^{AB}\label{eq:58}
\end{equation}
in the case of the usual, velocity-independent potentials. Then instead
of (\ref{eq:49}) we obtain 
\begin{equation}
\mathcal{T}=\mathcal{T}_{tr}+\mathcal{T}_{int}=\frac{1}{2m}g^{ij}p_{i}p_{j}
+\frac{1}{2}\left(J^{-1}\right)_{AB}p^{A}\!_{i}p^{B}\!_{j}g^{ij},\label{eq:59}
\end{equation}
where, obviously, $g^{ij}$ is the contravariant inverse of $g_{ij}$,
and $\left(J^{-1}\right)_{AB}$ is the covariant inverse of $J^{AB}$,
\begin{equation}
g^{ik}g_{kj}=\delta^{i}\!_{j},\quad\left(J^{-1}\right)_{AC}J^{CB}=
\delta_{A}\!^{B}.\label{eq:60}
\end{equation}
Obviously, if there is a velocity-dependence in the potential, then
the formula for $\mathcal{T}$ is more complicated. For example, the
presence of magnetic fields results in the configuration-dependent
translational gauging of canonical momenta. But in a moment we are
not interested in such details. Making use of the duality (\ref{eq:46})
we can write the Hamiltonian form of (\ref{eq:57}) as follows: 
\begin{equation}
\mathcal{T}_{int}=\frac{1}{2I}\underset{a}{\sum}P_{a}\!^{2}+\frac{1}{8I}
\underset{a,b}{\sum}\frac{\left(M^{a}\!_{b}\right)^{2}}{\left(Q^{a}
-Q^{b}\right)^{2}}+\frac{1}{8I}\underset{a,b}{\sum}\frac{\left(N^{a}\!_{b}
\right)^{2}}{\left(Q^{a}+Q^{b}\right)^{2}},\label{eq:61}
\end{equation}
where the quantities $M^{a}\!_{b}$, $N^{a}\!_{b}$ are given by 
\begin{equation}
M^{a}\!_{b}:=-\widehat{\rho}^{a}\!_{b}-\widehat{\tau}^{a}\!_{b},\quad N^{a}\!_{b}:=\widehat{\rho}^{a}\!_{b}-\widehat{\tau}^{a}\!_{b},\label{eq:62}
\end{equation}
$Q^{a}$ are given by (\ref{eq:32}), and $P_{a}$ are their conjugate
momenta, 
\begin{equation}
P_{a}=p_{a}\exp\left(-q^{a}\right).\label{eq:63}
\end{equation}
It is seen that in (\ref{eq:61}) one deals with a kind of ``diagonalization''
of the expression for $\mathcal{T}_{int}$.

Equations of affine motion may be derived on the basis of the variational
principle for the following Lagrangian: 
\begin{equation}
L=T-V(x,\varphi),\label{eq:64}
\end{equation}
where $T$ is given by (\ref{eq:49}), or in the Hamiltonian terms:
\begin{equation}
\frac{dF}{dt}=\left\{ F,H\right\}, \label{eq:65}
\end{equation}
where $H$ is the Hamiltonian corresponding to $L$, and $F$ runs
over a set of $2n(n+1)$ independent phase-space functions. But independently
of this variational framework, they may be derived in general, on
the basis of d'Alembert principle. According to this principle, equations
of affine motion are obtained from the general equation of motion
of the underlying system of material points by taking the monopole
and dipole moments of the balance laws for the linear momentum. The
points is that the original equations should be modified by introducing
the reactions responsible for maintaining of the constraints. The
reactions them-selves do not vanish, but their monopole and dipole
moments do so. Because of this, the effective, free of unspecified
reactions equations of affine motion have the form of the balance
laws for the total linear momentum and the total affine momentum (hypermomentum,
affine spin): 
\begin{equation}
\frac{dk^{i}}{dt}=F^{i},\quad\frac{dK^{ij}}{dt}=\frac{d\varphi^{i}\!_{A}}{dt}
\frac{d\varphi^{j}\!_{B}}{dt}J^{AB}+N^{ij}\label{eq:66}
\end{equation}
with the following meaning of symbols: 
\begin{eqnarray}
k&=&\int v(P)d\mu(P),\label{eq:67a}\\
K&=&\int\overrightarrow{\mathfrak{o}_{\phi}(P)\phi(P)}\otimes v(P)d\mu(P),
\label{eq:67b}\\
N&=&\int\overrightarrow{\mathfrak{o}_{\phi}(P)\phi(P)}\otimes F(P)d\mu(P).\label{eq:67c}
\end{eqnarray}
Therefore, $k$ is the total linear momentum, $K$ is the total dipole
moment of the momentum distribution (taken with respect to the instantaneous
position of the centre of mass) and $N$ is the total dipole momentum
(affine torque), also related to the centre of mass instantaneous
position. Reaction forces responsible for affine constraints are automatically
cancelled in (\ref{eq:67a})--(\ref{eq:67c}). The formulas (\ref{eq:67a})--(\ref{eq:67c}) are general, but their affine versions are just
\begin{equation}
k^{i}=m\frac{dx^{i}}{dt},\quad K^{ij}=\varphi^{i}\!_{A}\frac{d\varphi^{j}\!_{B}}{dt}J^{AB}\label{eq:68}
\end{equation}
in the sense of symbols (\ref{eq:2}).

Let us observe that the usual Legendre transformation (for velocity-indepen\-dent,
usual potentials) identifies those kinematical quantities with Hamiltonian
ones $p_{i}$, $\Sigma^{i}\!_{j}$ up to the index position, inessential
in Cartesian coordinates: 
\begin{equation}
p_{i}=g_{ij}k^{j},\quad\Sigma^{i}\!_{j}=K^{im}g_{mj}.\label{eq:69}
\end{equation}
It also convenient to use the co-moving representation of linear momentum, affine spin and affine moment of forces:
\begin{eqnarray}
\widehat{p}^{A}&=&\left.\left(\varphi^{-1}\right)^{A}\right._{i}k^{i},
\label{eq:70a}\\
\widehat{K}^{AB}&=&\left.\left(\varphi^{-1}\right)^{A}\right._{i}\left.
\left(\varphi^{-1}\right)^{B}\right._{j}K^{ij},\label{eq:70b}\\
\widehat{N}^{AB}&=&\left.\left(\varphi^{-1}\right)^{A}\right._{i}\left.
\left(\varphi^{-1}\right)^{B}\right._{j}N^{ij}.\label{eq:70c}
\end{eqnarray}

As it was said, the general, non-variational equations of motion are given
by (\ref{eq:66}), (\ref{eq:68}). Let us quote a few equivalent forms
also based on (\ref{eq:66}) with substituted (\ref{eq:68}), like,
e.g., 
\begin{equation}
\frac{dk^{i}}{dt}=F^{i},\quad\frac{dK^{ij}}{dt}=\Omega^{i}\!_{m}K^{mj}
+N^{ij}.\label{eq:71}
\end{equation}

Let us also observe that one can write: 
\begin{equation}
\frac{dK^{ij}}{dt}=N^{ij}+2\frac{\partial T_{int}}{\partial g_{ij}}.\label{eq:72}
\end{equation}
In particular, for the doubled skew-symmetric part of $K^{ij}$, i.e.,
for the angular momentum, we have 
\begin{equation}
\frac{dS^{ij}}{dt}=\frac{dK^{ij}}{dt}-\frac{dK^{ji}}{dt}=N^{ij}-N^{ji}=
\mathcal{N}^{ij},\label{eq:73}
\end{equation}
where the right-hand side denotes the usual torque. If $N^{ij}$ is
symmetric, in particular vanishing, this becomes the usual conservation
of angular momentum. The purely Lagrangian, i.e., $U$-based form
of equations of motion, may be formulated as follows: 
\begin{eqnarray}
\frac{d\widehat{k}^{A}}{dt}&=&-\widehat{k}^{B}\left(J^{-1}\right)_{BC}
\widehat{K}^{CA}+\widehat{F}^{A},\label{eq:74a}\\
\frac{d\widehat{K}^{AB}}{dt}&=&-\widehat{K}^{AC}\left(J^{-1}\right)_{CD}
\widehat{K}^{DB}+\widehat{N}^{AB}.\label{eq:74b}
\end{eqnarray}
Expressing these equations in terms of kinematical quantities one
obtains 
\begin{eqnarray}
m\frac{d\widehat{v}^{A}}{dt}&=&-m\widehat{\Omega}^{A}\!_{B}\widehat{v}^{B}
+\widehat{F}^{A},\label{eq:75a}\\
\frac{d\widehat{\Omega}^{B}\!_{C}}{dt}J^{CA}&=&
-\widehat{\Omega}^{B}\!_{D}\widehat{\Omega}^{D}\!_{C}J^{CA}
+\widehat{N}^{AB}.\label{eq:75b}
\end{eqnarray}
The explicit form of equations of motion reads:
\begin{eqnarray}
m\frac{d^{2}x^{i}}{dt^{2}} & = & F^{i}\left(x^{j},\frac{dx^{j}}{dt};\varphi^{k}\!_{A},
\frac{d\varphi^{k}\!_{A}}{dt};t\right),\label{eq:76a}\\
\varphi^{i}\!_{A}\frac{d^{2}\varphi^{j}\!_{B}}{dt^{2}}J^{AB} & = & N^{ij}\left(x^{m},\frac{dx^{m}}{dt};\varphi^{k}\!_{C},
\frac{d\varphi^{k}\!_{C}}{dt};t\right).\label{eq:76b}
\end{eqnarray}

The assumed non-singularity of matrices $\left[\varphi^{i}\!_{A}\right]$,
$\left[J^{AB}\right]$ in principle enables one to solve the second
equation with respect to second derivatives $d^{2}\varphi^{i}\!_{A}/dt^{2}$,
expressing them through dynamical variables. Nevertheless, the form
(\ref{eq:76b}) is more convenient, because it is geometrically suited
to the nature of our problem, in particular to additional constraints
which may be imposed on the affine motion. This follows from the fact
that in affinely-rigid behaviours the formula for the power of forces
is given by 
\begin{equation}
\mathcal{P}=\mathcal{P}_{tr}+\mathcal{P}_{int}=F^{j}g_{ij}v^{i}
+N^{jk}g_{ik}\Omega^{i}\!_{j}=F{}_{i}v^{i}+N^{j}\!_{i}\Omega^{i}\!_{j}.
\label{eq:77}
\end{equation}
And if for $F^{j}$, $N^{jk}$ one substitutes reactions maintaining
constraints, this expression vanishes; only the external given forces
contribute here. Similarly, if we subject the general affine motion
to some natural group-theoretical constraints, in a consequence of
which, e.g., $\Omega$ does belong to some Lie subalgebra of $L(V)$
or to some other linear subspace of clear algebraic meaning, then
the effective reaction-free system of equations of motion consists
of some natural subspace of (\ref{eq:76b}) and, obviously, of the explicit
description of constraints. This would not be the case if we used the
form of (\ref{eq:76b}) solved with respect to the second derivatives
of $\varphi^{j}\!_{B}$. Let us quote a few convincing examples of
both holonomic and non-holonomic constraints.

\subsection{Metrically rigid motion}

It consists in that both the mappings $\phi$, $\varphi$ are metrical
isometries, therefore, all distances and angles are preserved during
the motion, so that the following holds:
\begin{equation}
\eta_{AB}=g_{ij}\varphi^{i}\!_{A}\varphi^{j}\!_{B}.\label{eq:78}
\end{equation}
This explicit holonomic representation implies that $\Omega^{i}\!_{j}$, $\widehat{\Omega}^{A}\!_{B}$
are respectively $g$- and $\eta$-skew-symmetric during
any admissible motion, 
\begin{equation}
\Omega^{i}\!_{j}=-\Omega_{j}\!^{i}=-g_{ja}g^{ib}\Omega^{a}\!_{b},
\quad\widehat{\Omega}^{A}\!_{B}=-\widehat{\Omega}_{B}\!^{A}=
-\eta_{BC}\eta^{AD}\widehat{\Omega}^{C}\!_{D}.\label{eq:79}
\end{equation}
Therefore, (\ref{eq:76b}) is not valid any longer, instead the following
equations with unspecified reaction torques hold: 
\begin{equation}
\frac{dK^{ij}}{dt}=\frac{d\varphi^{i}\!_{A}}{dt}\frac{d\varphi^{j}\!_{B}}
{dt}J^{AB}+N^{ij}+\left.N_{R}\right.^{ij},\label{eq:80}
\end{equation}
or equivalently 
\begin{equation}
\varphi^{i}\!_{A}\frac{d^{2}\varphi^{j}\!_{B}}{dt^{2}}J^{AB}=N^{ij}
+\left.N_{R}\right.^{ij}.\label{eq:81}
\end{equation}
But the d'Alembert principle tells us that the power $\mathcal{P}_{R}$
of gyroscopic reactions vanishes on every $\Omega^{i}\!_{j}$ compatible
with constraints, i.e., $g$-skew-symmetric:
\begin{equation}
\mathcal{P}_{R}=\left.N_{R}\right.^{i}\!_{j}\Omega^{j}\!_{i}=
\left.N_{R}\right.^{ij}\Omega_{ji}=0\label{eq:82}
\end{equation}
if (\ref{eq:79}) holds. Therefore, 
\begin{equation}
\left.N_{R}\right.^{ij}=\left.N_{R}\right.^{ji},\quad\left.N_{R}
\right.^{i}\!_{j}=\left.N_{R}\right._{j}\!^{i}=g_{ja}g^{ib}\left.
N_{R}\right.^{a}\!_{b}.\label{eq:83}
\end{equation}
But this simply means that the reaction-free gyroscopic equations
of motion consist of the skew-symmetric part of (\ref{eq:80}) or
(\ref{eq:81}) with substituted (\ref{eq:78}), e.g., 
\begin{equation}
\varphi^{i}\!_{A}\frac{d^{2}\varphi^{j}\!_{B}}{dt^{2}}J^{AB}-
\varphi^{j}\!_{A}\frac{d^{2}\varphi^{i}\!_{B}}{dt^{2}}J^{AB}=N^{ij}-N^{ji}
=\mathcal{N}^{ij}.\label{eq:84}
\end{equation}
In other words, any parametrization of the isometry manifold (``Euler
angles'', ``rotation vectors'', etc.) may be safely substituted
to (\ref{eq:84}). The torque $\mathcal{N}^{ij}=N^{ij}-N^{ji}$ becomes
the function of those parameters and we obtain the system of $n\left(n-1\right)/2$
independent of equations (\ref{eq:84}) imposed on $n\left(n-1\right)/2$
parameters. It is clear that this system is the balance law for spin:
\begin{equation}
\frac{dS^{ij}}{dt}=\frac{d}{dt}\left(K^{ij}-K^{ji}\right)=\mathcal{N}^{ij}.\label{eq:85}
\end{equation}
It becomes the spin conservation when the torque $\mathcal{N}^{ij}$
does vanish, i.e., when $N^{ij}$ is symmetric. One can as well rewrite
(\ref{eq:85}) in Lagrangian terms:
\begin{eqnarray}
\frac{d\widehat{S}^{AB}}{dt}=\frac{d}{dt}\left(\widehat{K}^{AB}-
\widehat{K}^{BA}\right)&=&
\widehat{K}^{BC}\left(J^{-1}\right)_{CD}\widehat{K}^{DA}\nonumber\\
&-&\widehat{K}^{AC}\left(J^{-1}\right)_{CD}\widehat{K}^{DB}+
\widehat{\mathcal{N}}^{AB},\label{eq:86}
\end{eqnarray}
where 
\begin{equation}
\mathcal{\widehat{N}}^{AB}=\widehat{N}^{AB}-\widehat{N}^{BA}.\label{eq:87}
\end{equation}
This may be easily expressed in terms of co-moving affine velocity:
\begin{equation}
\frac{d\widehat{\Omega}^{B}\!_{C}}{dt}J^{CA}-\frac{d\widehat{\Omega}^{A}\!_{C}}{dt}J^{CB}=\widehat{\Omega}^{A}\!_{D}\widehat{\Omega}^{D}\!_{C}J^{CB}-\widehat{\Omega}^{B}\!_{D}\widehat{\Omega}^{D}\!_{C}J^{CA}+\widehat{\mathcal{N}}^{AB}.\label{eq:88}
\end{equation}
Those are Euler equations. When the co-moving inertial tensor is spherical,
i.e., when 
\begin{equation}
J=\frac{I}{2}\eta,\label{eq:89}
\end{equation}
then the non-dynamical terms on the right-hand side of (\ref{eq:88})
do vanish, and we obtain simply 
\begin{equation}
I\frac{d\widehat{\Omega}^{AB}}{dt}=I\frac{d\widehat{\Omega}^{A}\!_{C}}{dt}\eta^{CB}=\widehat{\mathcal{N}}^{AB}.\label{eq:90}
\end{equation}

\subsection{Shape-preserving motion}

Now the shape of the body is preserved, but not necessarily its size,
so that 
\begin{equation}
g_{ij}\varphi^{i}\!_{A}\varphi^{j}\!_{B}=\lambda\eta_{AB},\label{eq:91}
\end{equation}
where $\lambda$ denotes the time-dependent coefficient. Then, in
analogy to the previous example, the d'Alembert principle tells us
that the reactions-free equations of the constrained motion consist
of the skew-symmetric part and the trace of (\ref{eq:66}) or (\ref{eq:76b}):
\begin{eqnarray}
\frac{dS^{ij}}{dt}&=&\frac{d}{dt}\left(K^{ij}-K^{ji}\right)=\mathcal{N}^{ij}=
N^{ij}-N^{ji},\label{eq:92a}\\
\frac{dK^{i}\!_{i}}{dt}&=&\frac{d}{dt}\left(g_{ij}K^{ij}\right)=
g_{ij}\frac{d\varphi^{i}\!_{A}}{dt}\frac{d\varphi^{j}\!_{B}}{dt}J^{AB}
+g_{ij}N^{ij}=2T+N^{i}\!_{i},\qquad\label{eq:92b}
\end{eqnarray}
when written in a few independent forms. Using directly the representation
in terms of coordinates, we obtain that
\begin{eqnarray}
\varphi^{i}\!_{A}\frac{d^{2}\varphi^{j}\!_{B}}{dt^{2}}J^{AB}-\varphi^{j}\!_{A}
\frac{d^{2}\varphi^{i}\!_{B}}{dt^{2}}J^{AB}&=&\mathcal{N}^{ij}=N^{ij}-N^{ji},
\label{eq:93a} \\
g_{ij}\varphi^{i}\!_{A}\frac{d^{2}\varphi^{j}\!_{B}}{dt^{2}}J^{AB}&=&
g_{ij}N^{ij}.\label{eq:93b}
\end{eqnarray}

\subsection{Incompressible affine motion}

Incompressibility (isochoric motion) means that 
\begin{equation}
\frac{d}{dt}\det\left[\varphi^{i}\!_{A}\right]=0.\label{eq:94}
\end{equation}
This condition is well defined, although $\det\left[\varphi^{i}\!_{A}\right]$
is not a scalar but scalar density with respect to both spatial and
material coordinate transformations.

The identity 
\begin{equation}
\frac{d}{dt}\det\left[\varphi^{i}\!_{A}\right]=
\det\left[\varphi^{j}\!_{B}\right]\left(\varphi^{-1}\right)^{A}\!_{i}
\frac{d\varphi^{i}\!_{A}}{dt}=0\label{eq:95}
\end{equation}
is equivalent to 
\begin{equation}
Tr\,\Omega=Tr\,\widehat{\Omega}=\left(\varphi^{-1}\right)^{A}\!_{i}
\frac{d\varphi^{i}\!_{A}}{dt}=0.\label{eq:96}
\end{equation}
D'Alembert principle implies that reactions $N_{R}$ which keep these
constraints, being dual to the subspace of all trace-less matrices,
\begin{equation}
\mathcal{P}_{R}=\left.N_{R}\right.^{i}\!_{j}\Omega^{j}\!_{i}=0,\quad
\Omega^{k}\!_{k}=0,\label{eq:97}
\end{equation}
are proportional to the identity mapping, 
\begin{equation}
\left.N_{R}\right.^{i}\!_{j}=\lambda\delta^{i}\!_{j},\quad\left.N_{R}\right.^{ij}=\lambda g^{ij}.\label{eq:98}
\end{equation}
Therefore, the effective reactions-free system of equations of motion
is given by the trace-less part of the original balance law for $K^{ij}$,
i.e., explicitly 
\begin{equation}
\varphi^{i}\!_{A}\frac{d^{2}\varphi^{j}\!_{B}}{dt^{2}}J^{AB}-\frac{1}{n}g_{ab}\varphi^{a}\!_{A}\frac{d^{2}\varphi^{b}\!_{B}}{dt^{2}}J^{AB}g^{ij}=N^{ij}-\frac{1}{n}g_{ab}N^{ab}g^{ij}.\label{eq:99}
\end{equation}
This is a system of $\left(n^{2}-1\right)$ independent equations
of motion imposed on $\left(n^{2}-1\right)$ independent parameters
of $\varphi^{i}\!_{A}$ (cf. (\ref{eq:94})).

\subsection{Spatially rotation-less motion \label{point4}}

We have seen above that the concept of the purely rotational degrees
of freedom is well defined and correctly formulated: simply the mappings
$\phi$, $\varphi$ are isometries. It is not so with the opposite concept
of rotation-free, i.e., purely deformative configurations. The first,
naive idea would be to base this concept on the polar decomposition,
either in left or right version. So, rotation-free configurations
would be ones given by the purely deformative factor in the polar
decomposition. Therefore, depending on whether one deals with the
$U$-left or $U$-right version of (\ref{eq:35}), we would say that
$\varphi$ is purely deformative when it coincides with its $\eta$-
or $g$-symmetric part $A[\varphi]$ or $B[\varphi]$. However, this
would be incorrect. The more incorrect would be attempts of introducing
the pure deformation on the basis of the two-polar decomposition (\ref{eq:33}).
There are a few deep geometric reasons for that. First of all, the
symmetric mappings $A[\varphi]$, $B[\varphi]$ do not describe any
configurations at all. It is only mappings from $U$ to $V$ that
may be used as a model of the configuration, neither the linear automorphisms
of $U$ nor those of $V$. Without fixing some standard element of
the manifold of isometries $O\left(U,\eta;V,g\right)$ we cannot
identify automorphisms of the material or physical spaces with any
mappings of $U$ on to $V$. So, even from this relatively naive point
of view, the symmetric mappings do not describe configurations. But
there are also other arguments. Namely, even if we ``forget'' about
the above fact and simply proceed with the $\mathbb{R}^{n}$-model
of space and body manifolds, using the elements of $GL(n,\mathbb{R})$
as $LI(U,V)$, it is still so that the symmetric matrices do not form
a Lie group. Therefore, the corresponding absence of rotation is not an
equivalence relation because of the transitivity failure. If some
configurations $A$, $B$ are mutually non-rotated in the polar sense,
i.e., they are related by the symmetric matrix ${Sym}(A,B)$,
and if so are $B$, $C$ in the sense of being obtained from each other
by the action of some ${Sym}(B,C$), then in general $A$, $C$
are not connected by a symmetric matrix. It is well known that the
symmetric matrices do not form a Lie group or Lie algebra. The product
${Sym}(A,B)\,{Sym}(B,C)$ in general is not symmetric;
instead it splits into the multiplication of some symmetric matrix
and some nontrivial isometry. So, certainly, being related by a symmetric
matrix is not an equivalence relation, and the concept of mutually
rotation-free configurations is not correct. But there are well-defined
rotation-less motions. We say that the motion $\mathbb{R}\ni t\rightarrow\varphi(t)\in LI(U,V)$
is spatially rotation-less when $\Omega$ is $g$-symmetric:
\begin{equation}
\Omega^{i}\!_{j}-\Omega_{j}\!^{i}=\Omega^{i}\!_{j}-g_{jk}g^{il}
\Omega^{k}\!_{l}=0.\label{eq:100}
\end{equation}
And similarly, we say that it is materially rotation-less when $\widehat{\Omega}$
is $\eta$-symmetric: 
\begin{equation}
\widehat{\Omega}^{A}\!_{B}-\widehat{\Omega}_{B}\!^{A}=\widehat{\Omega}^{A}\!_{B}-\eta_{BC}\eta^{AD}\widehat{\Omega}^{C}\!_{D}=0.\label{eq:101}
\end{equation}
The symmetry of $\Omega$ or $\widehat{\Omega}$ is just the natural
complementary concept of their antisymmetry in rigid motion. And therefore,
this is a proper definition of the rotation-less behaviour, just behaviour,
not configuration. The point is that the symmetric matrices do not
form a Lie algebra. On the contrary, they are anti-Lie algebras in
the sense that their commutators are respectively $g$-
and $\eta$-skew-symmetric:
\begin{eqnarray}
\left[{Sym}\left(L\left(V\right),g\right),{Sym}\left(L\left(V\right),g\right)
\right]&=&{Asym}\left(L\left(V\right),g\right)\simeq SO\left(V,g\right)^{\prime},\label{eq:102a} \\
\left[{Sym}\left(L\left(U\right),\eta\right),{Sym}\left(L\left(U\right),
\eta\right)\right]&=&{Asym}\left(L\left(U\right),\eta\right)\simeq SO\left(U,\eta\right)^{\prime}.\qquad\label{eq:102b}
\end{eqnarray}
This is an interesting example of non-holonomic constraints, in
a sense different than the classical constraints of slide-free motion.
Nevertheless, some relationship with the usual non-holonomic
problems of non-sliding motion still seems to exist in certain hypothetical
applications. Let us consider, e.g., an affine motion of a small inclusion
or droplet suspension in very viscous fluid. It is natural to expect
that the surface friction may be an obstacle against rotations. And
then probably the effective constraints of rotation-less motion may
appear.

Let us stress some circumstance. Namely, the holonomic gyroscopic constraints
may be written alternatively in two apparently non-holonomic forms:
\begin{equation}
\Omega^{i}\!_{j}+\Omega_{j}\!^{i}=0,\quad\widehat{\Omega}^{A}\!_{B}+\widehat{\Omega}_{B}\!^{A}=0.\label{eq:103}
\end{equation}
They are mutually equivalent. On the other side, the two versions
of non-holonomic constraints (\ref{eq:100}) and (\ref{eq:101}) are non-equivalent. Namely, the $g$-symmetry of $\Omega$ is equivalent to the $\widehat{G}$-symmetry of $\widehat{\Omega}$ where, as usual, $\widehat{G}_{AB}$ denotes the Green deformation tensor, so that (\ref{eq:100}) is identical with
\begin{equation}
G_{AC}\widehat{\Omega}^{C}\!_{B}-G_{BC}\widehat{\Omega}^{C}\!_{A}=0.\label{eq:104}
\end{equation}

In a moment we are unable to answer the question concerning the details
of this relationship and the possible fields of physical applications.
From a perhaps naive point of view, it seems to be so that it is rather
the Euler symmetry (\ref{eq:100}) that seems to be applicable to
description of the affine motion of suspensions in viscous fluids. 

In any case, it is an interesting and rather new problem to discuss
the structure of equations of motion subject to rotation-less non-holonomic constraints. Again the d'Alembert principle shows the
advantage of the $K$-balance form of equations. Namely, the effective,
reactions-free equations are given by the symmetric part of the balance
laws, 
\begin{equation}
\varphi^{i}\!_{A}\frac{d^{2}\varphi^{j}\!_{B}}{dt^{2}}J^{AB}+\varphi^{j}\!_{A}
\frac{d^{2}\varphi^{i}\!_{B}}{dt^{2}}J^{AB}=N^{ij}+N^{ji},\label{eq:105}
\end{equation}
together with the algebraically substituted constraints (\ref{eq:100}).
The right-hand side of (\ref{eq:105}) depends only on given forces
and is free of reactions. The Lagrange form of (\ref{eq:105}) is given
by
\begin{equation}
\frac{d\widehat{\Omega}^{B}\!_{C}}{dt}J^{CA}+\frac{d\widehat{\Omega}^{A}
\!_{C}}{dt}J^{CB}=-\widehat{\Omega}^{B}\!_{D}\widehat{\Omega}^{D}\!_{C}J^{CA}
-\widehat{\Omega}^{A}\!_{D}\widehat{\Omega}^{D}\!_{C}J^{CB}+\widehat{N}^{AB}
+\widehat{N}^{BA},\label{eq:106}
\end{equation}
where $\widehat{\Omega}$ is subject to (\ref{eq:104}).

\subsection{Materially rotation-less motion\label{point5}}

It is also non-holonomic and somehow related to the spatially rotation-less
situation, nevertheless, in our opinion it is a bit less intuitive.
Now the material gyration is assumed to be $\eta$-symmetric, i.e.,
(\ref{eq:101}) is assumed to hold. The effective, reaction-free
equations of motion may be written as follows: 
\begin{eqnarray}
 &  & \frac{d\widehat{K}^{AC}}{dt}\widehat{\mathcal{D}}_{C}\!^{B}
+\frac{d\widehat{K}^{BC}}{dt}\widehat{\mathcal{D}}_{C}\!^{A}=
\widehat{N}^{AC}\widehat{\mathcal{D}}_{C}\!^{B}
+\widehat{N}^{BC}\widehat{\mathcal{D}}_{C}\!^{A}\nonumber\\
 &  & -\widehat{K}^{AM}\left(J^{-1}\right)_{MN}\widehat{K}^{NC}\widehat{\mathcal{D}}
_{C}\!^{B}-\widehat{K}^{BM}\left(J^{-1}\right)_{MN}\widehat{K}^{NC}
\widehat{\mathcal{D}}_{C}\!^{A},\label{eq:107}
\end{eqnarray}
where, obviously, $\widehat{K}^{AB}$ are co-moving components of
$K^{ij}$, and 
\begin{equation}
\widehat{K}^{AB}=\widehat{\Omega}^{B}\!_{C}J^{AC},\quad\widehat{\mathcal{D}}_{A}\!^{B}=G_{AC}\eta^{CB}.\label{eq:108}
\end{equation}
These equations are much more complicated than those for the spatially
rota\-tion-less motion. Namely, their non-dynamical terms depend on
the Green tensor, therefore, also on the configuration $\varphi$.
The Euler form is also complicated: 
\begin{equation}
\varphi^{i}\!_{A}\frac{d^{2}\varphi^{b}\!_{B}}{dt^{2}}J^{AB}g_{bc}C^{cj}
+\varphi^{j}\!_{A}\frac{d^{2}\varphi^{b}\!_{B}}{dt^{2}}J^{AB}g_{bc}C^{ci}=
N^{ib}g_{bc}C^{cj}+N^{jb}g_{bc}C^{ci},\label{eq:109}
\end{equation}
where 
\begin{equation}
C^{ab}=\varphi^{a}\!_{A}\varphi^{b}\!_{B}\eta^{AB}\label{eq:110}
\end{equation}
is the inverse Cauchy tensor.

\section{Dynamical symmetries of affine motion}

The non-holonomic constraints of rotation-less motion, i.e., the above
examples described in subsections \ref{point4} and \ref{point5}, are really exceptional and in a sense surprising within the realm of constrained affine motion. Let us stress to avoid some easy misunderstandings: they are really
non-holonomic and have nothing to do with apparently suggestive constraints
of the type that $\varphi$ in (\ref{eq:35}) is symmetric. Moreover,
our analysis above shows that such a formulation would be inconsistent,
just because of fixing some of infinitely possible isometries
$U[\varphi]$. And, let us repeat, the symmetric matrices do not form
a Lie group. The symmetry of $\Omega$ or $\widehat{\Omega}$ leads
to certain equations satisfied by $U[\varphi]$, $A[\varphi]$, $B[\varphi]$
in (\ref{eq:35}), but these equations are differential, not algebraic
ones. The $g$-symmetry of $\Omega$ or $\eta$-symmetry of $\widehat{\Omega}$
are the only natural counterparts of their antisymmetry in rigid motion.
And in any case, they are geometrically interesting special cases
of constraints, worth to be investigated from the very point of view
of purely analytical mechanics. 

It is interesting to ``solve'' the constraints equations (\ref{eq:100}),
i.e., to ``paramet\-rize'' somehow the manifold of non-holonomic constraints.
The best candidates are suggested by the polar decomposition (\ref{eq:35}).
Let us remind that $U[\varphi]$ is an isometry and that $A[\varphi]$
is $\eta$-symmetric, thus, 
\begin{eqnarray}
 &  & \eta_{AB}=g_{ij}U[\varphi]^{i}\!_{A}U[\varphi]^{j}\!_{B},\label{eq:111a}\\
 &  & \eta_{AC}A^{C}\!_{B}=\eta_{BC}A^{C}\!_{A},\quad\eta_{AC}\frac{d}{dt}A^{C}\!_{B}
=\eta_{BC}\frac{d}{dt}A^{C}\!_{A},\label{eq:111b}
\end{eqnarray}
and the co-moving angular velocity of the $U$-rotator, $\widehat{\omega}\in O\left(U,\eta\right)^{\prime}\subset L\left(U\right)$
is given by (\ref{eq:48}) and is, obviously, $\eta$-skew-symmetric:
\begin{equation}
\eta_{AC}\widehat{\omega}^{C}\!_{B}=-\eta_{BC}\widehat{\omega}^{C}\!_{A}.
\label{eq:112}
\end{equation}
Substituting those conditions to the definition (\ref{eq:19}) of the
affine velocity $\Omega$, we obtain after easy calculations the conclusion
that 
\begin{equation}
\widehat{\omega}^{A}\!_{B}=\frac{1}{2}\left(\left(A^{-1}\right)^{A}\!_{C}
\frac{dA^{C}\!_{B}}{dt}-\frac{dA^{A}\!_{C}}{dt}\left(A^{-1}\right)^{C}\!_{B}
\right).\label{eq:113}
\end{equation}
Therefore, the angular velocity of the $U$-rotator equals the half
of the commutator of two algebraically independent instantaneous quantities
$A^{-1}$, $dA/dt$: 
\begin{equation}
\widehat{\omega}=\frac{1}{2}\left[A^{-1},\frac{dA}{dt}\right].\label{eq:114}
\end{equation}
In any case, this quantity in general does not vanish and this reflects
the non-holonomic character of our constraints of non-rotational motion.
It is something different than the constancy of $U$, i.e., the vanishing
of $\widehat{\omega}$. Making use of the polar decomposition (\ref{eq:35})
and gyroscopic angular velocity (\ref{eq:48}), we can, a bit formally,
write down the constraints equations (\ref{eq:114}) in the following
Pfaff form: 
\begin{equation}
U^{-1}dU-\frac{1}{2}A^{-1}dA+\frac{1}{2}\left(dA\right)A^{-1}=0.\label{eq:115}
\end{equation}
This system of Pfaff equations is evidently non-integrable.

Before going any further with the analysis of the constrained affine motion,
let us quote a few remarks concerning the invariance problems. We
are interested mainly in the internal, i.e., relative, motion and
concentrate on the spatial and material rotational invariance. The
second, i.e., internal, equation (\ref{eq:76b}) is $O(V,g)$-invariant
if for any its solution $t\rightarrow\varphi(t)$ and for any $A\in O(V,g)$
the motion $t\rightarrow A\varphi(t)$ is also a solution. This implies
that 
\begin{equation}
N^{ij}\left(A\varphi,A\frac{d\varphi}{dt}\right)=A^{i}\!_{k}A^{j}\!_{l}N^{kl}
\left(\varphi,\frac{d\varphi}{dt}\right),\label{eq:116}
\end{equation}
where we do not indicate explicitly the possible explicit time-dependence
of $N$. But (\ref{eq:116}) means that the co-moving representation
of $N$ is non-sensitive with respect to the action of $A\in O(V,g)$:
\begin{equation}
\widehat{N}\left(A\varphi,A\frac{d\varphi}{dt}\right)=\widehat{N}\left(\varphi,
\frac{d\varphi}{dt}\right).\label{eq:117}
\end{equation}
This means that $\widehat{N}$ is algebraically built of the co-moving
quantities $\widehat{G}$, $\widehat{\Omega}$ and any fixed material
tensor $\widehat{K}$ in $\widehat{U}$: 
\begin{equation}
\widehat{N}\left(\varphi,\frac{d\varphi}{dt}\right)=
\widehat{F}\left(G,\widehat{\Omega},\widehat{K}\right).\label{eq:118}
\end{equation}
It is interesting that this form of $\widehat{N}$ implies the rotational
invariance of equations of motion, but it does not imply the conservation
of spin, i.e., internal angular momentum. Spin is conserved only if $\widehat{N}$
is a symmetric tensor, just like in mechanics of micropolar or micromorphic
continua. Let us mention in particular that the Green-Ostrogradskij
theorem implies that in orthonormal Cartesian coordinates the affine
momentum of forces is proportional to the mean value of Cauchy stress
tensor in the medium: 
\begin{equation}
N^{ij}=-\int\sigma^{ij}.\label{eq:119}
\end{equation}
In the case if hyperelastic bodies, both continuous and discrete,
$N^{ij}$, $\widehat{N}^{AB}$ are automatically symmetric. Indeed,
the condition for the potential energy
\begin{equation}
\mathcal{V}\left(A\varphi\right)=\mathcal{V}\left(\varphi\right),\quad A\in O\left(V,g\right),\label{eq:120}
\end{equation}
implies that $\mathcal{V}$ is algebraically built of the Green deformation
tensor: 
\begin{equation}
\mathcal{V}\left(\varphi\right)=W\left(G\left[\varphi\right],
\widehat{K}\right),\label{eq:121}
\end{equation}
where $K$ again denotes any state-independent tensor in $U$. And then
one can show immediately that 
\begin{equation}
\widehat{N}\left(\varphi\right)^{AB}=2\frac{\partial W}{\partial G_{AB}}=\widehat{N}\left(\varphi\right)^{BA},\quad N^{ij}=N^{ji},\label{eq:122}
\end{equation}
therefore, spin is a conserved quantity.

This was about the invariance under the left-hand side action of $O\left(V,g\right)$
on internal/relative degrees of freedom. Let us now ask what are conditions
of the invariance under the right-hand side action of material orthonormal
group $O\left(U,\eta\right)$. One can easily show that for any solution
$t\rightarrow\varphi(t)$ of (\ref{eq:76b}) and for any $B\in O(U,\eta)$
the right-rotated motion $t\rightarrow\varphi(t)B$ is a solution
too when the following holds: 
\begin{equation}
\varphi^{i}\!_{K}\frac{d^{2}\varphi^{j}\!_{L}}{dt^{2}}B^{K}\!_{C}B^{L}\!_{D}
J^{CD}=N^{ij}\left(\varphi B,\frac{d\varphi}{dt}B\right).\label{eq:123}
\end{equation}

Unlike in the case of spatial isotropy, this implies two conditions --- the internal and dynamical ones: 
\begin{equation}
J=I\eta,\quad N\left(\varphi B,\frac{d\varphi}{dt}B\right)=N\left(\varphi,\frac{d\varphi}{dt}\right).
\label{eq:124}
\end{equation}
The second conditions in (\ref{eq:124}) implies that $N$ depends on the mechanical state $\left(\varphi,d\varphi/dt\right)$ through the pair
$\left(C,\Omega\right)$ and any fixed, i.e., state-independent tensors
$K$ in $V$: 
\begin{equation}
N\left(\varphi,\frac{d\varphi}{dt}\right)=H\left(C,\Omega,K\right).
\label{eq:125}
\end{equation}

For hyperelastic bodies with the right-invariant potential energy,
\begin{equation}
\mathcal{V}\left(\varphi\right)=\mathcal{V}\left(\varphi B\right),\qquad B\in O\left(U,\eta\right),
\end{equation}
the following holds: 
\begin{equation}
\mathcal{V}\left(\varphi\right)=W\left(C\left[\varphi\right],K\right),
\label{eq:126}
\end{equation}
where again $K$ denotes any system of state-independent tensors in
$V$.

An important question appears as to when the dynamics of an affine
hyperelastic body is simultaneously isotropic in space and matter.
Obviously, this holds only when (\ref{eq:121}), (\ref{eq:124}), (\ref{eq:126})
are simultaneously satisfied. Therefore, the inertial tensor is spherical,
$J=I\eta$, and the potential energy $\mathcal{V}$ depends on $\varphi$
only through the deformation invariants, e.g., through the quantities
$I_{k}$ (\ref{eq:29}), or any of alternative expressions like $\lambda_{a}$, $q^{a}$, $Q^{a}$
or other used in the two-polar decomposition like (\ref{eq:32}).
Therefore, 
\begin{equation}
\mathcal{V}\left(\varphi\right)=F\left(I_{1},\dots,I_{n}\right)=
G\left(\lambda_{1},\dots,\lambda_{n}\right),\label{eq:127}
\end{equation}
where, obviously, $G$ is invariant under the group $S^{(n)}$ of all
permutations of its arguments.

It is clear that according to the general rules of Hamiltonian mechanics,
in the potential motion of the affinely-rigid body the reactions-free
affine moment of forces is given by 
\begin{equation}
N^{i}\!_{j}=-\varphi^{i}\!_{A}\frac{\partial\mathcal{V}}{\partial\varphi^{j}\!_{A}},\quad N^{ij}=-\varphi^{i}\!_{A}\frac{\partial\mathcal{V}}{\partial\varphi^{k}\!_{A}}g^{kj},\label{eq:128}
\end{equation}
and similar formulas hold for the co-moving representation.

The relationships (\ref{eq:70a})--(\ref{eq:70c}) imply that in a general, not necessarily hyperelastic, case equations of internal motion are simultaneously
spatially and materially isotropic, when $J=I\eta$ and $\widehat{N}$
is given by (\ref{eq:118}) with $\widehat{K}=\eta$ or, equivalently,
by (\ref{eq:125}) with $K=g$. For example, in a rather academic
elastic, but not necessarily hyperelastic, situation using the Cayley-Hamilton
theorem one can show that
\begin{equation}
\left.\widehat{N}_{el}\right.^{A}\!_{B}=\overset{n}{\underset{a=1}{\sum}}B_{a}
\left(I_{1},\dots,I_{n}\right)\left(\widehat{G}^{a-1}\right)^{A}\!_{B},
\label{eq:129}
\end{equation}
where $\left.\widehat{N}_{el}\right.^{A}\!_{B}$, $\widehat{G}^{A}\!_{B}$
are components of $\widehat{N}_{el}$, $\widehat{G}$ with the $\eta$-lowered
index $B$:
\begin{equation}
\left.\widehat{N}_{el}\right.^{A}\!_{B}=\left.\widehat{N}_{el}\right.^{AC}\eta_{CB},\quad\widehat{G}^{A}\!_{B}=\widehat{G}^{AC}\eta_{CB},\label{eq:130}
\end{equation}
and the scalar coefficients $B_{a}$ in expansion (\ref{eq:129})
depend on deformation invariants. One can show that in the hyperelastic
case, when the potential $\mathcal{V}\left(I_{1},\dots,I_{n}\right)$
does exist, the coefficients are given by the following derivatives: 
\begin{equation}
B_{a}=-2a\frac{\partial\mathcal{V}}{\partial I_{a}}.\label{eq:131}
\end{equation}

Obviously, the physical utility of elastic but not hyperelastic models
is rather doubtful, nevertheless it must be admitted for the completeness
of the theory.

Another example of a doubly isotropic model is one concerning the
isotropic internal friction in continuum droplet. The viscous stress
tensor is given in a linear approximation by 
\begin{equation}
\sigma_{visc}^{ij}=2\nu d^{ij}+\left(\zeta-\frac{2\nu}{n}\right)g_{ab}d^{ab}g^{ij},\label{eq:132}
\end{equation}
where the constants $\nu$, $\zeta$ are viscosity coefficients and
$d^{ij}$ is the deformation rate tensor. In the case of affine body
it is given by 
\begin{equation}
d^{ij}=\frac{1}{2}\left(\Omega^{ij}+\Omega^{ji}\right),\quad\Omega^{ij}=\Omega^{i}\!_{k}g^{kj}.\label{eq:133}
\end{equation}
Then, making use of the obvious formula (\ref{eq:119}) we obtain that
\begin{equation}
N_{visc}^{ij}=-V_{0}\sqrt{\frac{\det\left[g_{ij}\right]}{\det\left[\eta_{AB}
\right]}}\det\left[\varphi^{i}\!_{A}\right]\left(\nu\left(\Omega^{ij}
+\Omega^{ji}\right)+\left(\zeta-\frac{2\nu}{n}\right)\Omega^{k}\!_{k}g^{ij}
\right),\label{eq:134}
\end{equation}
where $V_{0}$ denotes the standard (Lagrangian) volume of the affine
body. This agrees with the formula (\ref{eq:125}) with $K=g$. 

The total viscoelastic and doubly-isotropic moment of forces is given
by 
\begin{equation}
N^{ij}=N_{el}^{ij}+N_{visc}^{ij}\label{eq:135}
\end{equation}
with the separate terms like (\ref{eq:129}) (\ref{eq:134}).

Those were interesting and instructive examples of the doubly isotropic
(spatially and materially) internal forces $N^{ij}$. It
is also interesting to find a description of more general isotropic
forces, adapted to certain special parametrizations of the configuration
space. In particular, some possibilities of partial separation of
variables, or rather their subsystems, appear then. First of all,
let us begin with the polar splitting (\ref{eq:35}), more precisely,
with its first form where the orthogonal term $U\left[\varphi\right]$
stands on the left-hand side. As mentioned above, gyroscopic kinetics
is described by the co-moving angular velocity $\widehat{\omega}=U^{-1}dU/dt$,
whereas deformation (together with its orientation with respect to
the body) is represented by the $\eta$-symmetric and positive factor
$A\left[\varphi\right]$ in (\ref{eq:35}). As usual, the following
tensors with $\eta$-shifted indices will be employed: 
\begin{eqnarray}
J^{A}\!_{B}=J^{AC}\eta_{CB},&\quad& \widehat{N}^{A}\!_{B}=\widehat{N}^{AC}\eta_{CB},\label{eq:136a}\\ G^{A}\!_{B}=\eta^{AC}G_{CB},&\quad& A^{KL}=A^{K}\!_{M}\eta^{ML}.
\label{eq:136b}
\end{eqnarray}
It is clear that 
\begin{equation}
\widehat{\Omega}=A^{-1}\widehat{\omega}A+A^{-1}\frac{dA}{dt}=
A^{-1}\left(\widehat{\omega}+\frac{dA}{dt}A^{-1}\right)A\label{eq:137}
\end{equation}
and 
\begin{equation}
N^{ij}=\varphi^{i}\!_{C}\varphi^{j}\!_{D}\widehat{N}^{CD}=
U^{i}\!_{K}U^{j}\!_{L}A^{K}\!_{C}A^{L}\!_{D}\widehat{N}^{CD}.\label{eq:138}
\end{equation}
This suggests us to introduce the following quantity: 
\begin{equation}
\overline{N}^{KL}=A^{K}\!_{C}A^{L}\!_{D}\widehat{N}^{CD},\qquad\textrm{i.e.},
\qquad\overline{N}=\left(A\otimes A\right)\widehat{N}.\label{eq:139}
\end{equation}
Just like $\widehat{N}$ itself, $\overline{N}$ is also an element
of $U\otimes U$, however of a quite different nature. Namely, $\widehat{N}^{AB}$
are components of $N$ with respect to the basis $\varphi E_{A}$, $A=1,\dots,n$,
affinely co-moving with the body. Unlike this, the quantities $\overline{N}^{AB}$
are components of $N$ with respect to the orthonormal basis $U\left[\varphi\right]E_{A}$, $A=1,\dots,n$,
co-moving with the $U\left[\varphi\right]$-gyroscope of the polar
decomposition of $\varphi$.

After this substitution, our internal equations of motion, i.e., the
second subsystem (\ref{eq:76b}), become as follows: 
\begin{eqnarray}
A^{K}\!_{C}J^{C}\!_{D}\frac{d^{2}A^{DM}}{dt^{2}}-A^{K}\!_{C}J^{C}\!_{D}A^{D}
\!_{E}\frac{d\widehat{\omega}^{EM}}{dt}
-2A^{K}\!_{C}J^{C}\!_{D}\frac{dA^{D}\!_{E}}{dt}\widehat{\omega}^{EM}&&
\nonumber\\
+A^{K}\!_{C}J^{C}\!_{D}A^{D}\!_{E}\widehat{\omega}^{E}\!_{F}
\widehat{\omega}^{FM} =\overline{N}^{KM}&&\label{eq:140}
\end{eqnarray}
with the convention (\ref{eq:136a})--(\ref{eq:136b}) concerning the $\eta$-shift of tensor indices. Obviously, for the spatially isotropic models one
is faced with some kind of partial separation of variables. Indeed,
the spatial isotropy means that $\overline{N}$ is independent on
the variable $U$. It is a function of the state quantities 
$A$, $dA/dt$, $\widehat{\omega}$ only. Roughly speaking, in the non-holonomic $\widehat{\omega}$-representation it is a kind of cyclic state variable. Therefore, the procedure of solving equations of motion splits into three steps:
\begin{enumerate}
\item[1)] (\ref{eq:140}) is a system of differential equations for the time
dependence of quantities $A$, $\widehat{\omega}$. 

\item[2)] Assuming that the previous step is done, we write the system of
differential equations for $U$,
\begin{equation}
\frac{dU}{dt}=U\widehat{\omega}\left(t\right).\label{eq:141}
\end{equation}
Let us observe that this system is time-dependent through the time
evolution of $\widehat{\omega}\left(t\right)$.

\item[3)] When the steps 1), 2) are performed, we construct the final solution:
\begin{equation}
\varphi\left(t\right)=U\left(t\right)A\left(t\right).\label{eq:142}
\end{equation}
\end{enumerate}

Obviously, this is only the general scheme. For dynamically realistic
models the steps 1), 2) as a rule, are not analytically solvable.
Nevertheless, even this partial separation and a sequence of procedures
is very helpful for the understanding the problem. In any case, its
structure looks simpler and more adapted to operations. Let us only
quote two examples corresponding to (\ref{eq:129}), (\ref{eq:134}).
After substituting (\ref{eq:136a})--(\ref{eq:136b}) we find respectively that
\begin{eqnarray}
\overline{N}_{el}&=&\overset{n}{\underset{a=1}{\sum}}B_{a}\left(I_{1}\dots I_{n}\right)A^{2a},\label{eq:143}\\
\overline{N}_{visc}&=&-V_{0}\det A\left[\nu
\left(\frac{dA}{dt}A^{-1}+A^{-1}\frac{dA}{dt}\right)\right.\nonumber\\
&+&\left.\left(\zeta-\frac{2\nu}{n}\right)
Tr\left(\frac{dA}{dt}A^{-1}\right)\eta^{-1}\right].\label{eq:144}
\end{eqnarray}
It is a nice feature of the both formulas that $\overline{N}_{el}$
depends only on $A$, and $\overline{N}_{visc}$ depends only on $A$, $dA/dt$;
there is no dependence on $\widehat{\omega}$. This independence is
due to the fact that $\overline{N}_{visc}$ describes the internal friction.
To be honest, the linear dependence of $\overline{N}_{visc}$ on the
$g$-symmetric part of $\Omega$ is an approximation valid in the
case of small internal velocities. In general, the higher powers of
$\Omega^{(ij)}$ are admissible.

Both expressions (\ref{eq:129}), (\ref{eq:134}), therefore, also
(\ref{eq:143}), (\ref{eq:144}), have an additional interesting feature
of being isotropic simultaneously in space and material. It
is natural to ask for the optimal way of expressing this fact. As
expected, the most natural way consists in using the two-polar representation
(\ref{eq:33}) and the related quantities (\ref{eq:32}), (\ref{eq:44a}), (\ref{eq:44b}), (\ref{eq:45a}), (\ref{eq:45b}).
Then, identifying the factors $L$, $R$ in (\ref{eq:33}) with linear
mappings from $\mathbb{R}^{n}$ to $V$ and $U$ respectively, and
similarly identifying $\widehat{\chi}$, $\widehat{\vartheta}$, $D$
with linear mappings from $\mathbb{R}^{n}$ to $\mathbb{R}^{n}$,
we obtain the following formulas: 
\begin{eqnarray}
\Omega & = & L\left(\widehat{\chi}+\frac{dD}{dt}D^{-1}-D\widehat{\vartheta}D^{-1}\right)
L^{-1},\label{eq:145a}\\
\widehat{\Omega} & = & R\left(D^{-1}\widehat{\chi}D+D^{-1}\frac{dD}{dt}-\widehat{\vartheta}\right)
R^{-1},\label{eq:145b}\\
\widehat{\omega} & = & R\left(\widehat{\chi}-\widehat{\vartheta}\right)R^{-1},
\label{eq:145c}
\end{eqnarray}
obviously $\widehat{\omega}$ (\ref{eq:48}) is a linear mapping
from $U$ to U. These formulas are simple and suggestive. The last
of them, i.e., one for $\widehat{\omega}$, is an infinitesimal expression
of the obvious fact that $U=LR^{-1}$.

As mentioned in (\ref{eq:69}), for the potential systems
the Legendre transformation relates $\Sigma^{i}\!_{j}$ to $K^{i}\!_{j}=K^{im}g_{mj}$.
Similarly $\widehat{\Sigma}^{A}\!_{B}$ is related to $\widehat{K}^{AC}G_{CB}$.
It is important that the second index is lowered with the help of
Green deformation tensor, not with the help of the fixed material
metric $\eta_{CB}$. One should not confuse $\widehat{\Sigma}^{A}\!_{B}$
with 
\begin{equation}
\widehat{K}^{AB}=\left.\left(\varphi^{-1}\right)^{A}\!\right._{i}\left.
\left(\varphi^{-1}\right)^{B}\!\right._{j}K^{ij}.\label{eq:146}
\end{equation}
Let us quote the explicit Legendre formulas for those quantities in
the special materially isotropic case $J^{AB}=I\eta^{AB}$. So, we
have that
\begin{eqnarray}
\Sigma=K&=&IL\left(D\widehat{\vartheta}D+D\frac{dD}{dt}-D^{2}\widehat{\chi}
\right)L^{-1},\label{eq:147}\\
\widehat{\Sigma}=\varphi^{-1}K\varphi&=&IR\left(\widehat{\vartheta}D^{2}
+\frac{dD}{dt}D-D\widehat{\chi}D\right)R^{-1},\label{eq:148}\\
\widehat{K}=\varphi^{-1}K\left(\varphi^{-1}\right)^{T}&=&
IR\left(\widehat{\vartheta}+\frac{dD}{dt}D^{-1}-D\widehat{\chi}D\right)R^{-1}.
\label{eq:149}
\end{eqnarray}
Let us mention that the corresponding spin parts, i.e., doubled skew-symmetric
parts of those quantities, equal respectively to
\begin{eqnarray}
S=\Sigma-\Sigma^{T}&=&IL\left(2D\widehat{\vartheta}D-D^{2}\widehat{\chi}
-\widehat{\chi}D^{2}\right)L^{-1},\label{eq:150}\\
V=\widehat{\Sigma}^{T}-\widehat{\Sigma}&=&IR\left(2D\widehat{\chi}D
-D^{2}\widehat{\vartheta}-\widehat{\vartheta}D^{2}\right)R^{-1},\label{eq:151}\\
\widehat{S}=\widehat{K}-\widehat{K}^{T}&=&IR\left(-D\widehat{\chi}D^{-1}
-D^{-1}\widehat{\chi}D+2\widehat{\vartheta}\right)R^{-1}.\label{eq:152}
\end{eqnarray}
Then for the doubly (spatially and materially) isotropic
problems the quantities $S$, $V$ are constants of motion. Unlike
this, $\widehat{S}$, i.e., (\ref{eq:152}), is not a conserved quantity.

In (\ref{eq:139}) we have introduced the quantity $\overline{N}\in U\otimes U$,
the components of which represented $N$ with respect to the moving
orthonormal basis $U[\varphi]E_{A}$, $A=1,\dots,n$. This representation
enabled one to reduce equations of motion to the $U$-independent
form (\ref{eq:140}). Something similar may be done for the two-polar
representation. Namely, there exist an obvious analogy between (\ref{eq:137})
and (\ref{eq:145b}) in that 
\begin{equation}
\Omega=L\widetilde{\Omega}L^{-1},\quad
\widehat{\Omega}=R\underset{\widetilde{\;}}{\Omega}R^{-1},\quad
\widetilde{\Omega}=D\underset{\widetilde{\;}}{\Omega}D^{-1}.\label{eq:153-4}
\end{equation}
It is clear that the matrix elements of the $\mathbb{R}^{n}$-tensors,
\begin{equation}
\widetilde{\Omega}=\widehat{\chi}+\frac{dD}{dt}D^{-1}-
D\widehat{\vartheta}D^{-1},\quad\underset{\widetilde{\;}}{\Omega}=
D^{-1}\widehat{\chi}D+D^{-1}\frac{dD}{dt}-\widehat{\vartheta},\label{eq:155}
\end{equation}
are components of $\Omega$ with respect to the orthonormal
frame $L_{a}$ frozen into the Cauchy gyroscope and the components
of $\widehat{\Omega}$ with respect to the orthonormal frame co-moving
with the Green deformation tensor. And the same representation may
be introduced for any other tensor quantity, in particular for the
affine moment of forces $N$. The mixed, contravariant-covariant representation
of $\widetilde{N}$ is given by $\widetilde{N}^{a}\!_{b}$, where $\widetilde{N}=L^{-1}NL$, i.e.,
\begin{eqnarray}
 \widetilde{N}^{a}\!_{b}=\left\langle L^{a},NL_{b}\right\rangle =L^{a}\!_{i}N^{i}\!_{j}L^{j}\!_{b},\qquad
  N^{i}\!_{j}=N^{ik}g_{kj}.\label{eq:156}
\end{eqnarray}
Substituting the above equations to the doubly isotropic case of the
internal subsystem (\ref{eq:76b}), we obtain the following equations
of motion:
\begin{eqnarray} 
&&D\frac{d^{2}D}{dt^{2}}-D^{2}\frac{d\widehat{\chi}}{dt}
+D\frac{d\widetilde{\vartheta}}{dt}D-2D\frac{dD}{dt}\widehat{\chi}
+2D\widehat{\vartheta}\frac{dD}{dt}\nonumber\\
&&+D^{2}\widehat{\chi}^{2}-2D\widehat{\chi}D\widehat{\vartheta}
+D\widehat{\vartheta}^{2}D=\frac{1}{I}\widetilde{N}\left(D,\frac{dD}{dt},
\widehat{\chi},\widehat{\vartheta}\right).\label{eq:157}
\end{eqnarray}
The dynamical double isotropy implies that $\widetilde{N}$ depends
only on the indicated variables 
$D$, $dD/dt$, $\widehat{\chi}$, $\widehat{\vartheta}$
but is independent of the angular variables $L$, $R$. Therefore,
similarly like in (\ref{eq:140}), there is a partial separability
of the problem (\ref{eq:157}):
\begin{enumerate}
\item[1)] Just as it was the case with (\ref{eq:140}), one solves the system
(\ref{eq:157}). To be more precise, one dreams about solving this
system of $n^{2}$ ordinary differential equations for the $n^{2}$
dynamical variables $D$, $\widehat{\chi}$, $\widehat{\vartheta}$. Some
kind of rigorous solutions is possible only for the two-dimensional
case $n=2$. For higher dimensions, including the physical case $n=3$,
only some special solutions may be analytically found. 

\item[2)] When the time dependence $\mathbb{R}\ni t\rightarrow\left(\widehat{\chi}(t),\widehat{\vartheta}(t)\right)$
is ``known'', we substitute it to the definition of angular velocities:
\begin{equation}
\frac{dL}{dt}=L\widehat{\chi},\qquad\frac{dR}{dt}=\widehat{\vartheta}.
\label{eq:158}
\end{equation}
Then one obtains the system of differential equations with right-hand sides
explicitly dependent on time.

\item[3)] After ``solving'' (\ref{eq:158}) we substitute everything to (\ref{eq:33})
and obtain the final solution.
\end{enumerate}
As mentioned many times above, this partial reduction (separability)
of (\ref{eq:140}), (\ref{eq:157}) is rather ideal and qualitative,
nevertheless, it is helpful in understanding the dynamical structure
of spatially and doubly isotropic models.

\section{D'Alembert and vakonomic models of rotation-less motion}

Let us now discuss briefly the interesting special case of non-holonomic
rotation-less constraints. This will be rather an introductory analysis;
up to our knowledge nobody discussed this kind of constraints, either
in the d'Alembert or vakonomic version. From some point of view the
apparently exotic vakonomic form is rather simpler and more elegant
{[}6{]}. It is yet rather too early to try deciding which is more
physical and in what kind of problems.

Let us substitute formally the polar representation (\ref{eq:113}),
(\ref{eq:114}) of (\ref{eq:100}) to the polar expression of the
kinetic energy (\ref{eq:52}). Then we obtain that
\begin{eqnarray}
T_{int}&=&\frac{1}{8}\eta_{KL}\frac{dA^{K}\!_{A}}{dt}\frac{dA^{L}\!_{B}}{dt}
^{AB}+\frac{1}{4}\eta_{KL}\left(A^{-1}\right)^{K}\!_{D}\frac{dA^{D}\!_{C}}
{dt}A^{C}\!_{A}\frac{dA^{L}\!_{B}}{dt}J^{AB}\nonumber\\
&+&\frac{1}{8}\eta_{KL}\left(A^{-1}\right)^{K}\!_{E}\frac{dA^{E}\!_{C}}{dt}
A^{C}\!_{A}\left(A^{-1}\right)^{L}\!_{F}\frac{dA^{F}\!_{D}}{dt}A^{D}\!_{B}
J^{AB}.\label{eq:159}
\end{eqnarray}
After calculations this may be expressed in the following more concise
form: 
\begin{eqnarray}
T_{int}&=& \frac{1}{8}\eta_{KL}\left(A^{-1}\right)^{K}\!_{E}\left(A^{-1}\right)^{L}\!_{F}
\left(A^{E}\!_{C}\frac{dA^{C}\!_{A}}{dt}\right.\nonumber \\
&+&\left.\frac{dA^{E}\!_{C}}{dt}A^{C}\!_{A}\right) \left(A^{F}\!_{D}\frac{dA^{D}\!_{B}}{dt}+\frac{dA^{F}\!_{D}}{dt}A^{D}\!_{B}
\right)J^{AB}.\label{eq:160}
\end{eqnarray}
The variational derivative of $T_{int}$ with respect to the symmetric
tensor 
\begin{equation}
A_{AB}=\eta_{AC}A^{C}\!_{B}=A_{BA}\label{eq:161}
\end{equation}
is given by
\begin{eqnarray}
\left.\frac{\delta T_{int}}{\delta A_{AB}}\right|_{symm} & = & -\frac{1}{4}\frac{d^{2}}{dt^{2}}A^{(A}\!_{L}J^{B)L}-\frac{1}{4}\frac{d}{dt}
\left(\left(A^{-1}\right)\!^{(A}\!_{E}J^{B)L}\frac{dA^{E}\!_{C}}{dt}A^{C}
\!_{L}\right)\nonumber \\
 & - & \frac{1}{4}\eta_{KL}\frac{d}{dt}\left(\frac{dA^{K}\!_{E}}{dt}
\left(A^{-1}\right)\!^{L(A}A^{B)}\!_{D}\right)J^{ED}\nonumber \\
 & - & \frac{1}{4}\eta_{KL}\frac{d}{dt}\left(\left(A^{-1}\right)^{K}\!_{E}
\frac{dA^{E}\!_{C}}{dt}A^{C}\!_{F}\left(A^{-1}\right)\!^{L(A}A^{B)}\!_{D}
\right)J^{FD}\nonumber \\
 & - & \frac{1}{4}\eta_{KL}\frac{dA^{K}\!_{E}}{dt}\frac{dA^{F}\!_{D}}{dt}A^{D}\!_{G}
\left(A^{-1}\right)\!^{L(A}A^{B)}\!_{F}J^{EG}\nonumber\\
 & - & \frac{1}{4}\eta_{KL}\left(A^{-1}\right)^{K}\!_{E}\frac{dA^{E}\!_{C}}{dt}A^{C}
\!_{M}\frac{dA^{F}\!_{D}}{dt}A^{D}\!_{N}\left(A^{-1}\right)\!^{L(A}A^{B)}\!_{F}
J^{MN}\nonumber \\
 & + & \frac{1}{4}\eta_{KL}\frac{dA^{K}\!_{D}}{dt}\left(A^{-1}\right)^{L}\!_{E}
\frac{dA^{E(A}}{dt}J^{B)D}\nonumber \\
 & + & \frac{1}{4}\eta_{KL}\left(A^{-1}\right)^{K}\!_{E}\frac{dA^{E}\!_{C}}{dt}A^{C}
\!_{D}\left(A^{-1}\right)^{L}\!_{F}\frac{dA^{F(A}}{dt}J^{B)D}.\label{eq:162}
\end{eqnarray}

When there are hyperelastic forces derivable from the potential $\mathcal{V}$
depending only on the Green deformation tensor, then equations of
motion have the following form: 
\begin{equation}
\left.\frac{\delta T_{int}}{\delta A_{AB}}\right|_{symm}=-A_{KC}\eta^{K(A}\widehat{N}^{B)C},\label{eq:163}
\end{equation}
where 
\begin{equation}
\widehat{N}^{BC}=-\left(D\mathcal{V}\right)^{BC}.\label{eq:164}
\end{equation}

In spite of their apparently complicated structure, equations (\ref{eq:162})
are readable. And having them solved for the time dependence of $A_{AB}$,
we obtain from (\ref{eq:113})/(\ref{eq:114}) the time dependence of
$\widehat{\omega}$, and then, solving (in principle) (\ref{eq:48})
for dependence $t\rightarrow U(t)$, we finally obtain (in principle)
$\varphi=UA$.

Let us mention that all tensor indices are shifted from their natural
position with the help of $\eta$.

The usual d'Alembert procedure, i.e., the symmetric part of (\ref{eq:140})
with algebraically substituted constraints (\ref{eq:100}), i.e.,
(\ref{eq:113})/(\ref{eq:114}), leads to the following form, less readable
than (\ref{eq:162}), (\ref{eq:163}), (\ref{eq:164}):
\begin{eqnarray}
&&J^{AB}\frac{d^{2}A^{B(C}}{dt^{2}}A^{D)}\!_{A}- J^{A}\!_{B}A^{B}\!_{E}\frac{d}{dt}\frac{1}{2}\left(\left(A^{-1}\right)^{E}
\!_{F}\frac{d}{dt}\left(A^{F(C}\right)A^{D)}\!_{A}\right.\nonumber\\
&&\left.-\frac{d}{dt}\left(A^{E}
\!_{F}\right)\left(A^{-1}\right)^{F(C}A^{D)}\!_{A}\right)
-J^{A}\!_{B}\frac{dA^{B}\!_{E}}{dt}\left(\left(A^{-1}\right)^{E}\!_{F}
\frac{d}{dt}\left(A^{F(C}\right)A^{D)}\!_{A}\right.\nonumber\\
&&\left.-\frac{d}{dt}\left(A^{E}\!_{F}
\right)\left(A^{-1}\right)^{F(C}A^{D)}\!_{A}\right)
+\frac{1}{4}J^{A}\!_{B}A^{B}\!_{E}\left(\left(A^{-1}\right)^{E}\!_{G}
\frac{d}{dt}\left(A^{G}\!_{F}\right)\right.\nonumber\\
&&\left.-\frac{d}{dt}\left(A^{E}\!_{G}\right)
\left(A^{-1}\right)^{G}\!_{F}\right)
\left(\left(A^{-1}\right)^{F}\!_{H}\frac{d}{dt}\left(A^{H(C}\right)A^{D)A}
\right.\nonumber\\
&&\left.
-\frac{d}{dt}\left(A^{F}\!_{H}\right)\left(A^{-1}\right)^{H(C}A^{D)}\!_{A}
\right)\quad=\quad\overline{N}^{(CD)}.\label{eq:165}
\end{eqnarray}
Again the $\eta$-shift of indices is meant here. The difference between
(\ref{eq:162})/(\ref{eq:163}) and (\ref{eq:165}) on their right-hand
side is not essential, because it is only due to the $A$-term transformation
of $\widehat{N}$ into $\overline{N}$. They may be written in a similar
form in this sense. But the difference between other terms of (\ref{eq:164})
and (\ref{eq:165}) is more essential. The detailed analysis of this
difference is postponed to the next paper. In any case, it is a general
rule that the d'Alembert and vaconomic procedures give different equations.

\section*{Appendix: d'Alembert vs. vaconomic constraints}

The problem appear more than century ago. It is well known that when
the holonomic constraints 
\begin{equation}
F_{a}(q)=0,\quad a=1,\ldots,m,
\end{equation}
are imposed onto the motion of a Lagrangian dynamical system with
generalized coordinates $q^{1},\dots,q^{n}$, then one can equivalently
use the d'Alembert procedure or the restricted extremum (more precisely,
stationary value) problem. If Lagrangian is given by $L(q,\dot{q})$,
then the Lusternik theorem tells us that the conditional extremum
(more precisely, stationary value) 
\begin{equation}
\delta\int Ldt=0,\quad F_{a}(q)=0,\quad a=1,\ldots,m,
\end{equation}
is given by the functions of time satisfying equations:
\begin{equation}
\frac{d}{dt}\frac{\partial L}{\partial\dot{q}^{i}}-\frac{\partial L}{\partial q^{i}}=R_{i},\quad F_{a}(q)=0,
\end{equation}
where
\begin{equation}
R_{i}=\frac{d\mu^{a}}{dt}\omega_{ai},\qquad
\omega_{ai}=\frac{\partial F_{a}}{\partial q^{i}},\qquad 
\frac{dF_{a}}{dt}=\omega_{ai}(q)\frac{dq^{i}}{dt}.
\end{equation}
And those are exactly d'Alembert equations with the multipliers
\begin{equation}
\lambda^{a}=\frac{d\mu^{a}}{dt}.
\end{equation}
The same formulas for reaction forces $R_{i}$ hold also for non-variational,
e.g., dissipative dynamical models:
\begin{equation}
\frac{d}{dt}\frac{\partial L}{\partial\dot{q}^{i}}-\frac{\partial L}{\partial q^{i}}=Q_{i},
\end{equation}
where $Q_{i}$ are non-variational generalized forces. Then as well
we have that
\begin{equation}
\frac{d}{dt}\frac{\partial L}{\partial\dot{q}^{i}}-\frac{\partial L}{\partial q^{i}}=Q_{i}+\lambda^{a}\frac{\partial F_{a}}{\partial q^{i}},\quad F_{a}(q)=0.
\end{equation}
So, there was a natural temptation to expect something similar for
systems with non-holonomic constraints, for simplicity linear in velocities, 
\begin{equation}
\omega_{ai}(q)\frac{dq^{i}}{dt}=0,
\end{equation}
but without the intergrability assumption 
\begin{equation}
\omega_{ai}\frac{\partial F_{a}}{\partial q^{i}}=0,
\end{equation}
i.e., without the vanishing of exterior differentials:
\begin{equation}
\frac{\partial\omega_{ai}}{\partial q^{j}}-\frac{\partial\omega_{aj}}{\partial q^{i}}\neq 0
\end{equation}
But it turned out in contrary: d'Alembert procedure gives again the
equations
\begin{equation}
\frac{d}{dt}\frac{\partial L}{\partial\dot{q}^{i}}-\frac{\partial L}{\partial q^{i}}=Q_{i}+\lambda^{a}\omega_{ai},\quad\omega_{ai}(q)\frac{dq^{i}}{dt}=0,
\end{equation}
with reactions coefficients $\lambda^{a}$ to be eliminated. But the
Lusternik theorem for 
\begin{equation}
\delta\int Ldt=0,\qquad\omega_{ai}(q)\frac{dq^{i}}{dt}=0
\end{equation}
gives something drastically else: 
\begin{eqnarray}
\frac{d}{dt}\frac{\partial L}{\partial\dot{q}^{i}}-\frac{\partial L}{\partial q^{i}}=\frac{d\mu^{a}}{dt}\omega_{ai}+\mu^{a}\left(\frac{\partial\omega_{ai}}
{\partial q^{j}}-\frac{\partial\omega_{aj}}{\partial q^{i}}\right)\dot{q}^{j}&=&0,\\
\omega_{ai}(q)\frac{dq^{i}}{dt}&=&0.
\end{eqnarray}
The difference is obvious. Moreover, $\mu^{a}$ become a kind of dynamical
variables, because they occur both by itself and their time
derivatives. The problems of sliding-free rolling motion are ruled
by the d'Alembert procedure. But, on the other hand, the Lusternik variational,
i.e., vaconomic, procedure looks very interesting and intriguing. It
gives rice to the new mathematical discipline and its applications seem to be also possible, first of all, in active control problems. 

Our equations for the rolling-free affine motion in the d'Alembert
and vaconomic sense are also drastically different, although as yet
we are unable to express them in qualitative terms.

\section*{Acknowledgements}

This paper partially contains results obtained within the framework of the research project N N501 049 540 financed from the Scientific Research Support Fund in 2011-2014. The authors are greatly indebted to the Polish Ministry of Science and Higher Education for this financial support. The second and third authors (BG and VK) wish to thank the first author, professor Jan J. S\l{}awianowski, for inspiring and encouraging them to work on this subject.

\end{document}